%% file: desy21-130-with-supplement.tex
\documentclass[preprintnumbers,aps,superscriptaddress,notitlepage,nofootinbib]{revtex4-2}
\usepackage[english]{babel}
\usepackage[utf8]{inputenc}
\usepackage[colorinlistoftodos, color=green!40, prependcaption]{todonotes}
\usepackage{amsthm}
\usepackage{mathtools}
\usepackage{physics}
\usepackage{xcolor}
\usepackage{graphicx}
\usepackage[left=23mm,right=13mm,top=35mm,columnsep=15pt]{geometry}
\usepackage{adjustbox}                                                         
\usepackage{placeins}
\usepackage[T1]{fontenc}
\usepackage{lipsum}
\usepackage{csquotes}

\newcommand{\ptjet}{\ensuremath{p_{\mathrm{T}}^{\mathrm{jet}}}}

\newcommand{\etajetlab}{\ensuremath{\eta_{\mathrm{lab}}^{\mathrm{jet}}}}

\usepackage[colorlinks=true, urlcolor=blue,citecolor=blue,anchorcolor=blue]{hyperref}

\newcommand{\sectionPRL}[1]{ \textbf{ #1.}}

\begin{document}
\preprint{DESY 21-130, ISSN 0418-9833}
\title{Measurement of lepton-jet correlation in deep-inelastic scattering \\ with the H1 detector using machine learning for unfolding}
\input{desy21-130.H1authorlist.revtexnothanks}
\date{\today}
\begin{abstract}
  The first measurement of lepton-jet momentum imbalance and azimuthal correlation in lepton-proton scattering at high momentum transfer is presented. These data, taken with the H1 detector at HERA, are corrected for detector effects using an unbinned machine learning algorithm (\textsc{MultiFold}), which considers eight observables simultaneously in this first application. The unfolded cross sections are compared to calculations performed within the context of collinear or transverse-momentum-dependent (TMD) factorization in Quantum Chromodynamics (QCD) as well as Monte Carlo event generators.
  {\em Accepted by PRL (Feb 25, 2022)}. 
\end{abstract}

\maketitle
\sectionPRL{Introduction}
Studies of jets produced in high energy scattering experiments have played a crucial role in establishing Quantum Chromodynamics (QCD) as the fundamental theory underlying the strong nuclear force~\cite{Ali:2010tw}.
During the current era of the Large Hadron Collider (LHC), experimental, theoretical, and statistical advances have ushered in a new era of precision QCD studies with jets~\cite{Salam:2009jx,Rabbertz:2017ssq} and their substructure~\cite{Larkoski:2017jix,Asquith:2018igt}.

\begin{figure}[ht]
    \centering
    \includegraphics[width=0.4\textwidth,trim=5 5 5 5,clip]{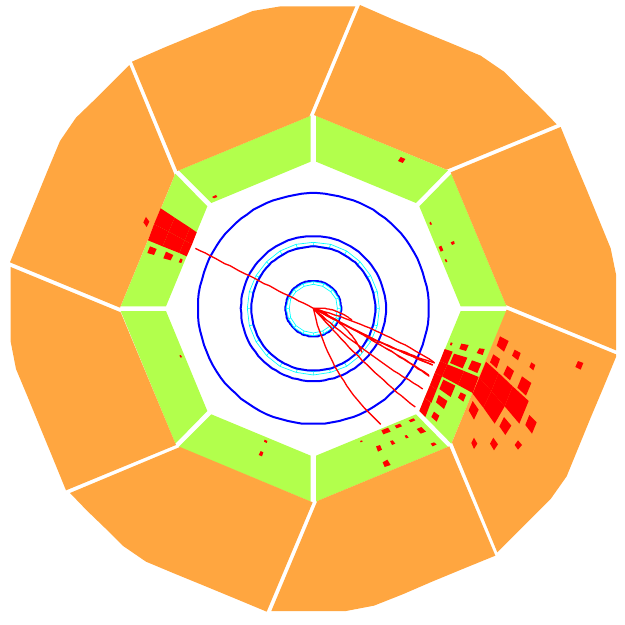} %
    \caption{A display of the H1 tracker and calorimeter detectors, showing a DIS event with
approximate Born kinematics, $eq\to eq$, which yields a lepton and a jet in a back-to-back topology perpendicular to the beam axis.}
    \label{fig:signature}
\end{figure}

These innovations motivate new measurements of hadronic final states in the deep inelastic scattering (DIS), $e+p\to e+X$, at the HERA collider.  DIS measurements provide high precision to study jets, because of the minimal backgrounds from the $ep$ initial state and the excellent segmentation, energy resolution, and calibration of the HERA experiments.  
In the DIS Born level limit, a virtual photon is exchanged with a quark inside the proton to create a back-to-back topology between the lepton and the resulting jet(s) as shown in Fig.~\ref{fig:signature}.  The Born level limit represented a background for most jet measurements by H1~\cite{Andreev:2016tgi, Adloff:1998vc, Adloff:2000tq, Adloff:2002ew, Aktas:2003ja, Aktas:2004px, Aktas:2007aa, Aaron:2009vs, Aaron:2010ac, Aaron:2011ef,Andreev:2014wwa} and ZEUS~\cite{Breitweg:2000sv, Chekanov:2001fw, Chekanov:2002be, Chekanov:2004hz, Chekanov:2006xr, Chekanov:2006yc, Abramowicz:2010cka, Abramowicz:2010ke}, which targeted higher-order QCD processes and were carried out in the Breit frame~\cite{Newman:2013ada}. 
While the one jet final state has been studied inclusively in terms of the scattered lepton kinematics to determine proton structure functions~\cite{Andreev:2013vha,Collaboration:2010ry,Aaron:2012qi,Chekanov:2009na,Aaron:2008ad}, the immense potential of the jet kinematics in this channel is only now being
realized. 

For example, single jet production has been proposed as a key channel for extracting quark transverse-momentum-dependent (TMD) parton distribution functions (PDFs)~\cite{Gutierrez-Reyes:2018qez,Gutierrez-Reyes:2019vbx,Liu:2018trl,Liu:2020dct,Kang:2020fka,Arratia:2020ssx}. In particular, measurements of back-to-back lepton-jet production $e+p\to e+{\rm jet}+X$ measured in the laboratory frame provide sensitivity to TMD PDFs in the limit when the imbalance $q_\mathrm{T}^\text{jet}=|\vec p_\mathrm{T}^{\, e}+\vec p_\mathrm{T}^{\, {\rm jet}}|$ of the transverse momentum of the scattered lepton ($p_\mathrm{T}^{\, e}$) and the jet ($p_\mathrm{T}^{\, {\rm jet}}$)  is relatively small ($q_\mathrm{T}^\text{jet}\ll p_\mathrm{T}^{\, e}\sim p_\mathrm{T}^{\, {\rm jet}}$)~\cite{Liu:2018trl}.  This corresponds to a small deviation from $\pi$ in azimuthal angle between the lepton and jet axes ($\Delta\phi^\text{jet}\equiv|\pi-(\phi^e-\phi^\text{jet})|$) in the transverse plane.
TMD PDFs are an essential ingredient for the quantum tomography of the proton that probes the origin of its spin, mass, size, and other properties. %

The energy dependence of TMD PDFs can also probe unexplored aspects of QCD as they follow %
a more complex set of evolution equations than collinear PDFs~\cite{Gribov:1972ri,Dokshitzer:1977sg,Altarelli:1977zs}, involving components that cannot be calculated using perturbation theory. A complete description remains open in part because of a lack of precise measurements over a wide kinematic range. Existing constraints from DIS data are at very low momentum transfer ($Q^2\approx$ 1 GeV$^{2}$) from fixed-target experiments~\cite{Aghasyan:2017ctw,Ashman:1991cj,Airapetian:2012ki,Adolph:2013stb,Avakian:2019drf}.  Drell-Yan production in fixed target~\cite{Dove:2021ejl, Baldit:1994jk,Falciano:1986wk,Conway:1989fs,Aghasyan:2017jop} and collider experiments~\cite{Abbott:1999wk, Abazov:2007ac, Abazov:2010kn, Affolder:1999jh, Aaltonen:2012fi, Aidala:2018ajl, Aaij:2016mgv, Aaij:2015zlq, Aaij:2015gna, Khachatryan:2016nbe, Chatrchyan:2011wt, Aad:2015auj, Aad:2014xaa} can provide TMD-sensitive measurements up to high scales ($Q^2\approx$ 10000 GeV$^{2}$). %
The HERA experiments can cover the entire kinematic region $Q^{2}\approx 1-10000$ GeV$^{2}$ so they can yield a key ingredient to connecting the existing experimental and theoretical information, including with lattice QCD calculations, which have made significant advances in describing aspects of TMD evolution~\cite{Ebert:2018gzl,Shanahan:2020zxr}.

This Letter presents a measurement of jet production in neutral current (NC) DIS events close to the Born level configuration, $eq\rightarrow eq$. The cross section of this process is measured differentially as a function of the jet transverse momentum and pseudorapidity, as well as lepton-jet momentum imbalance and azimuthal angle correlation. 
This measurement probes a range of QCD phenomena, including TMD PDFs and their evolution with energy. 
A novel machine learning (ML) technique called  \textsc{MultiFold}~\cite{Andreassen:2019cjw,Andreassen:2021zzk} is used to correct for detector effects for the first time in any experiment, enabling the simultaneous and unbinned unfolding of the target observables.

\sectionPRL{Experimental method}
The H1 detector\footnote{This measurement uses a right handed coordinate system defined such that the positive $z$ direction points in the direction of the proton beam and the nominal interaction point is located at $z=0$. The polar angle $\theta$, is defined with respect to this axis. The pseudorapidity is defined as $\eta_{\mathrm{lab}} = -\ln \tan(\theta/2)$.}~\cite{Abt:1993wz,Andrieu:1993kh,Abt:1996hi,Abt:1996xv,Appuhn:1996na} is a general purpose particle detector with cylindrical geometry. The main sub-detectors used in this analysis are the inner tracking detectors and the Liquid Argon (LAr) calorimeter, which are both immersed in a magnetic field of 1.16 T provided by a superconducting solenoid. The central tracking system, which covers 15$^{\circ}$ $<\theta<$ 165$^{\circ}$ and the full azimuthal angle, consists of drift and proportional chambers that are complemented with a silicon vertex detector in the range $30^{\circ}<\theta<150^{\circ}$~\cite{Pitzl:2000wz}. It yields a transverse momentum resolution for charged particles of $\sigma_{p_\mathrm{T}}/p_\mathrm{T}$ = 0.2$\%$ $p_\mathrm{T}$/GeV$~\oplus~$1.5$\%$. The LAr calorimeter, which covers $4^{\circ}<\theta< 154^{\circ}$ and full azimuthal angle, consists of an electromagnetic section made of lead absorbers and a hadronic section with steel absorbers; both are highly segmented in the transverse and longitudinal directions. Its energy resolution is $\sigma_{E}/E = 11\%/\sqrt{E/\mathrm{GeV}}$~$\oplus$~$1\%$ for leptons~\cite{Andrieu:1994yn} and $\sigma_{E}/E\approx 50\%/\sqrt{E/\mathrm{GeV}}$~$\oplus$~$3\%$ for charged pions~\cite{Andrieu:1993tz}.  In the backward region ($153^\circ < \theta < 177.5^\circ$), energies are measured with a lead-scintillating fiber calorimeter~\cite{H1SPACALGroup:1996ziw}.

 This offline analysis uses data collected with the H1 detector in the years 2006 and 2007 when positrons and protons were collided at energies of 27.6~GeV and 920~GeV, respectively. The total integrated luminosity of this data sample corresponds to 136~pb$^{-1}$~\cite{Aaron:2012kn}. 

This analysis follows an event selection used previously~\cite{Andreev:2014wwa}. The trigger used to select events requires a high energy cluster in the electromagnetic part of the LAr calorimeter. The scattered lepton is identified with the highest transverse momentum LAr cluster matched to a track, and is required to pass certain isolation criteria~\cite{Adloff:2003uh}. After fiducial cuts, the trigger efficiency is higher than 99.5$\%$~\cite{Aaron:2012qi,Andreev:2014wwa} for scattered lepton
candidates with energy $E_{e'}>11$ GeV. A series of fiducial and quality cuts based on simulations~\cite{Andreev:2014wwa,Andreev:2016tgi} suppress backgrounds to a negligible level. 

The kinematics of the DIS reaction can be described by the following variables: the square of the four-momentum transfer, $Q^{2}$, which sets the scale at which the proton is probed, and the inelasticity of the reaction, $y$, which is related to the scattering angle in the lepton-quark center-of-mass frame. The $\Sigma$ method~\cite{Bassler:1994uq} is used to reconstruct $y$ and $Q^{2}$ as: 
\begin{eqnarray*}
y &=& \frac{\sum_{i\in \mathrm{had}}(E_{i}-p_{i,z})}{\sum_{i\in \mathrm{had}}(E_{i}-p_{i,z})+E_{e^{'}}(1-\cos\theta_{e^{'}})}\\
Q^{2} &=&\frac{E_{e^{'}}^2 \sin^{2}\theta_{e^{'}} }{1-y}\,\\
\label{kinematicreco}
\end{eqnarray*}
where $\theta_{e^{'}}$ is the polar angle of the scattered lepton and $\sum(E_{i}-p_{i,z})$ is the total difference between the energy and longitudinal momentum of the entire hadronic final state (HFS).  After removing tracks and clusters associated to the scattered lepton, an energy flow algorithm~\cite{energyflowthesis,energyflowthesis2,energyflowthesis3} is used to define the HFS objects that enter the sum $\sum_{i\in \text{had}}$. Compared to other methods, the $\Sigma$ reconstruction reduces sensitivity to collinear initial state Quantum Electrodynamic (QED)
radiation, $e\to e\gamma$, since the beam energies are not included in the calculation. Events are required to have $45 <\sum(E_{i}-p_{i,z})< 65$~GeV to suppress initial-state QED radiation. Final state QED radiation is corrected for in the unfolding procedure. Correction factors to account for virtual and real higher-order QED effects are estimated using the simulations described below. Electroweak effects cancel in the normalized cross-sections to below the percent level and are neglected.  Events with $Q^{2}> 150$ GeV$^{2}$ and $0.08<y<0.7$ are selected for further analysis. 

Monte Carlo (MC) simulations are used to correct the data for detector acceptance and resolution effects. Two generators are used for this purpose: \textsc{Djangoh}~\cite{Charchula:1994kf}~1.4 and \textsc{Rapgap}~\cite{Jung:1993gf}~3.1. Both generators implement Born level matrix elements for the NC DIS, boson–gluon fusion, and QCD Compton processes and are interfaced with \textsc{Heracles}~\cite{Spiesberger:237380,Kwiatkowski:1990cx,Kwiatkowski:1990es} for QED radiation.  The CTEQ6L PDF set~\cite{Pumplin:2002vw} and the Lund hadronization model~\cite{Andersson:1983ia} with parameters fitted by the ALEPH Collaboration~\cite{Schael:2004ux} are used for the non-perturbative components.  \textsc{Djangoh} uses the Colour Dipole Model as implemented in \textsc{Ariadne}~\cite{Lonnblad:1992tz} for higher order emissions, and \textsc{Rapgap} uses parton showers in the leading logarithmic approximation. Each of these generators is combined with a detailed simulation of the H1 detector response based on the \textsc{Geant}3 simulation program~\cite{Brun:1987ma} and reconstructed in the same way as data.

The \textsc{FastJet}~3.3.2 package~\cite{Cacciari:2011ma,Cacciari:2005hq} is used to cluster jets in the laboratory frame with
the longitudinally-invariant, inclusive $k_{\mathrm{T}}$ algorithm~\cite{Catani:1993hr,Ellis:1993tq} and distance parameter $R = 1$. The inputs for the jet clustering are HFS objects with $-1.5<\eta_{\mathrm{lab}}<2.75$. Jets with transverse momentum $\ptjet>$ 5 GeV are selected for further analysis. %

The input for the jet clustering at the generator level (``particle level'') are final-state particles with proper lifetime $c\tau > 10$~mm generated with \textsc{Rapgap} or \textsc{Djangoh}, excluding the scattered lepton. Reconstructed jets are matched to the generated jets with an angular distance selection of $
\Delta R = \sqrt{(\phi_{\rm gen}^{\rm jet}- \phi_{\rm reco}^{\rm jet})^{2} + (\eta_{\rm gen}^{\rm jet}- \eta_{\rm reco}^{\rm jet})^{2}}  <0.9$.

The final measurement is presented in a fiducial volume defined by $Q^2>150$ GeV$^2$, $0.2 < y < 0.7$, $\ptjet>$ 10 GeV, and $-1.0<\eta^{\rm jet}_{\mathrm{lab}}<2.5$; the total inclusive jet cross section in this region is denoted $\sigma_\text{jet}$.

\sectionPRL{Unfolding method}
Following successful applications of artificial neural networks (NNs) to H1 event reconstruction~\cite{Kogler:2011zz,electronID,Andreev:2014wwa} the ML-based \textsc{MultiFold} technique~\cite{Andreassen:2019cjw,Andreassen:2021zzk} is used to correct for detector effects.  Unlike other widely used forms of unfolding based on regularized matrix inversion~\cite{DAgostini:1994fjx,Hocker:1995kb,Schmitt:2012kp}, \textsc{MultiFold} allows the data to be unfolded unbinned and simultaneously in many dimensions, due to the structure and flexibility of NNs.  Furthermore, unlike other approaches to unbinned~\cite{Aslan:2003vu,Lindemann:1995ut,Datta:2018mwd,Bellagente:2020piv,Bellagente:2019uyp,pmlr-v130-vandegar21a} or ML-based~\cite{Gagunashvili:2010zw,Glazov:2017vni,Bellagente:2020piv,Bellagente:2019uyp,Datta:2018mwd,pmlr-v130-vandegar21a} unfolding, \textsc{MultiFold} reduces to the widely studied iterative unfolding approach~\cite{1974AJ.....79..745L,Richardson:72,DAgostini:1994fjx} when the inputs are binned.  At each iteration, \textsc{MultiFold} employs NN classifiers to estimate likelihood ratios that are used as event weights.  At each iteration, a classifier is trained to distinguish data from simulation and then the corresponding weights at detector-level are inherited by the corresponding particle-level events in simulation.  To accommodate the stochastic nature of the detector response, a second classifier is used to distinguish the original simulation from the one with detector-level weights.  This produces a weighting map that is a proper function of the particle-level phase space.  The weights can then be applied to detector-level.  This process is repeated a total of five times.  The number of iterations is chosen such that the closure tests described below do not dominate the total uncertainty.  A brief technical review of the \textsc{MultiFold} method can be found in the Supplement, including the statistical origin of the reweighting~\cite{hastie01statisticallearning,sugiyama_suzuki_kanamori_2012} and properties of the neural networks \cite{pmlr-v9-glorot10a}.

The unfolding is performed simultaneously for eight observables ($\vec{p}_\mathrm{T}^e$, $p_z^e$, $p_\mathrm{T}^\text{jet}$, $\eta^\text{jet}$, $\phi^\text{jet}$, $q_\mathrm{T}^\text{jet}/Q$, and $\Delta\phi^\text{jet}$) and is unbinned.  The distributions of the four target observables ($p_\mathrm{T}^\text{jet}$, $\eta^\text{jet}$, $q_\mathrm{T}^\text{jet}/Q$, and $\Delta\phi^\text{jet}$) are presented as separate histograms for the quantitative comparison of predictions to data; the other observables provide a comprehensive set of possible migrations and detector effects of the target observables.  All NNs are implemented in \textsc{Keras}~\cite{keras} and \textsc{TensorFlow}~\cite{tensorflow} using the \textsc{Adam}~\cite{adam} optimization algorithm.  The networks have three hidden layers with 50, 100, and 50 nodes per layer, respectively, using rectified linear unit activation functions for intermediate layers and a sigmoid function for the final layer.  At each iteration/step, the data and simulations are split into 50\% for training, 50\% for validation, and all simulated events are used for the final results.  Binary cross-entropy is used as the loss function and training proceeds until the validation loss does not improve for 10 epochs in a row.  All of the algorithm hyperparameters are near their default values, with small changes made to qualitatively improve the precision across observables.

The statistical uncertainty of the measurement is determined using the bootstrap technique\footnote{For a discussion of the interplay between deep learning and the bootstrap, see e.g.~\cite{nixon2020why,austern2021asymptotics}.}~\cite{10.1214/aos/1176344552}. In particular, the unfolding procedure is repeated on 100 pseudo datasets, each constructed by resampling the data with replacement.  As the number of MC events significantly exceeds the number of data events, the MC dataset is kept fixed.  The resulting statistical uncertainty ranges from about 0.5 to 10$\%$ for the jet transverse momentum measurement, and it ranges from  0.5 to 3.5$\%$ for the other measurements. Variations from the random nature of the network initialization and training are found to be negligible compared to the data statistical uncertainty.

\sectionPRL{Uncertainties} Systematic uncertainties are evaluated by varying an aspect of the simulation and repeating the unfolding.  The procedures used here closely follow other recent H1 analyses~\cite{Andreev:2014wwa,Andreev:2016tgi}.
 The HFS-object energy scale uncertainty originates from two contributions: HFS objects contained in high $p_{\mathrm{T}}$ jets and other HFS objects. In both cases, the energy-scale uncertainty is $\pm$1$\%$~\cite{Andreev:2014wwa,Kogler:2011zz}. Both uncertainties are estimated separately by varying the corresponding HFS energy by $\pm1\%$.
    The uncertainty of the measurement of the azimuthal angle of the HFS objects is $\pm 20$ mrad.
    The uncertainty of the measurement of the energy of the scattered lepton ranges from $\pm0.5\%$ at backward and central regions~\cite{Aaron:2011gp} to $\pm$1$\%$ at forward regions~\cite{Andreev:2014wwa}. 
      The uncertainty of the measurement of the azimuthal angle of the scattered lepton is $\pm1$ mrad~\cite{Aaron:2012qi}. 
     The uncertainty associated with the modeling of the hadronic final state in the event generator used for unfolding and acceptance corrections is estimated by the difference between the results obtained using \textsc{Djangoh} and \textsc{Rapgap}. 
     Given that the differential cross sections are reported normalized to the inclusive jet cross section, normalization uncertainties such as luminosity scale or trigger efficiency cancel in the ratio.

The bias of the unfolding procedure is determined by taking the difference in the result when unfolding with \textsc{Rapgap} and with \textsc{Djangoh}.  This procedure gives a consistent result to unfolding detector-level \textsc{Rapgap} with \textsc{Djangoh} (and vice versa). It was also verified that unfolding \textsc{Rapgap} with itself using statistically independent samples gives unbiased results within MC statistical uncertainties.  The \textsc{Rapgap} and \textsc{Djangoh} distributions bracket the data and have rather different underlying models. Therefore, comparing the results with both generators provides a realistic evaluation of the procedure bias.  This uncertainty is typically below a few percent, but reaches 10\% at low $q_{\rm T}^\text{jet}/Q$. 

The total systematic uncertainty ranges from 2 to 25$\%$ for $\ptjet$; from 3 to 7$\%$ for $\eta^{\rm jet}_{\mathrm{lab}}$; from 4 to 15$\%$ in $q_{\rm T}^\text{jet}/Q$; and from 4 to 6$\%$ for $\Delta\phi^\text{jet}$.

\sectionPRL{Theory predictions} The unfolded data are compared to fixed order calculations within perturbative QCD (pQCD) and calculations within the TMD factorization framework. The pQCD calculation at next-to-next-to-leading order (NNLO)
accuracy in QCD (up to $\mathcal{O}(\alpha_s^2$)) was obtained with the \textsc{Poldis} code~\cite{Borsa:2020ulb,Borsa:2020yxh}, which is based on the Projection to Born Method~\cite{Cacciari:2015jma}. These calculations are multiplied by hadronization corrections that are obtained with \textsc{Pythia}~8.3~\cite{Sjostrand:2006za,Sjostrand:2014zea} using its default set of parameters. These corrections are smaller than 10$\%$ for most kinematic intervals and are consistent with corrections derived by an alternative generator, \textsc{Herwig}~7.2~\cite{Bellm:2015jjp,Bahr:2008pv}, using its default parameters.  The uncertainty of the calculations is given by the variation the factorization and renormalization scale $Q^2$ by a factor of two~\cite{Borsa:2020ulb,Borsa:2020yxh} as well as NLOPDF4LHC15 variations~\cite{Butterworth:2015oua}.
    
The TMD calculation uses the framework developed in Refs.~\cite{Liu:2018trl,Liu:2020dct} using the same jet radius and algorithm used in this work\footnote{This differs from the original paper~\cite{Liu:2018trl} using the anti-$k_{T}$ algorithm. The difference is power suppressed at the accuracy of the calculation.}. The inputs are TMD PDFs and soft functions derived in Ref.~\cite{Su:2014wpa}, which were extracted from an analysis of semi-inclusive DIS and Drell-Yan data. %
The calculation is performed at the next-to-leading logarithmic accuracy.  This calculation is performed within TMD factorization and no matching to the high $q_\mathrm{T}$ region is included, where the TMD approach is expected to be inaccurate. In contrast to pQCD calculations, the TMD calculations do not require non-perturbative corrections, because such effects are already included. Calculations with the TMD framework are available for the TMD sensitive cross sections, which are $q_\mathrm{T}^\text{jet}/Q$ and $\Delta\phi^\text{jet}$. Uncertainties are not yet available for the TMD predictions\footnote{The scale variation procedure that is standard in the collinear framework does not translate easily to the TMD framework \cite{Zhongbo2021}.}.  Additional TMD-based calculations are provided by the MC generator \textsc{Cascade}~\cite{Baranov:2021uol}, using matrix elements from \textsc{KaTie}~\cite{vanHameren:2016kkz} and parton branching TMD PDFs~\cite{Hautmann:2017xtx,Hautmann:2017fcj,Martinez:2018jxt}. A first setup integrates to HERAPDF2.0~\cite{Abramowicz:2015mha} and a second setup uses angular ordering and $p_{\mathrm{T}}$ as the renormalization scale~\cite{Martinez:2019mwt,Martinez:2020fzs}.

\begin{figure}[ht] 
    \centering
    \includegraphics[width=0.24\textwidth]{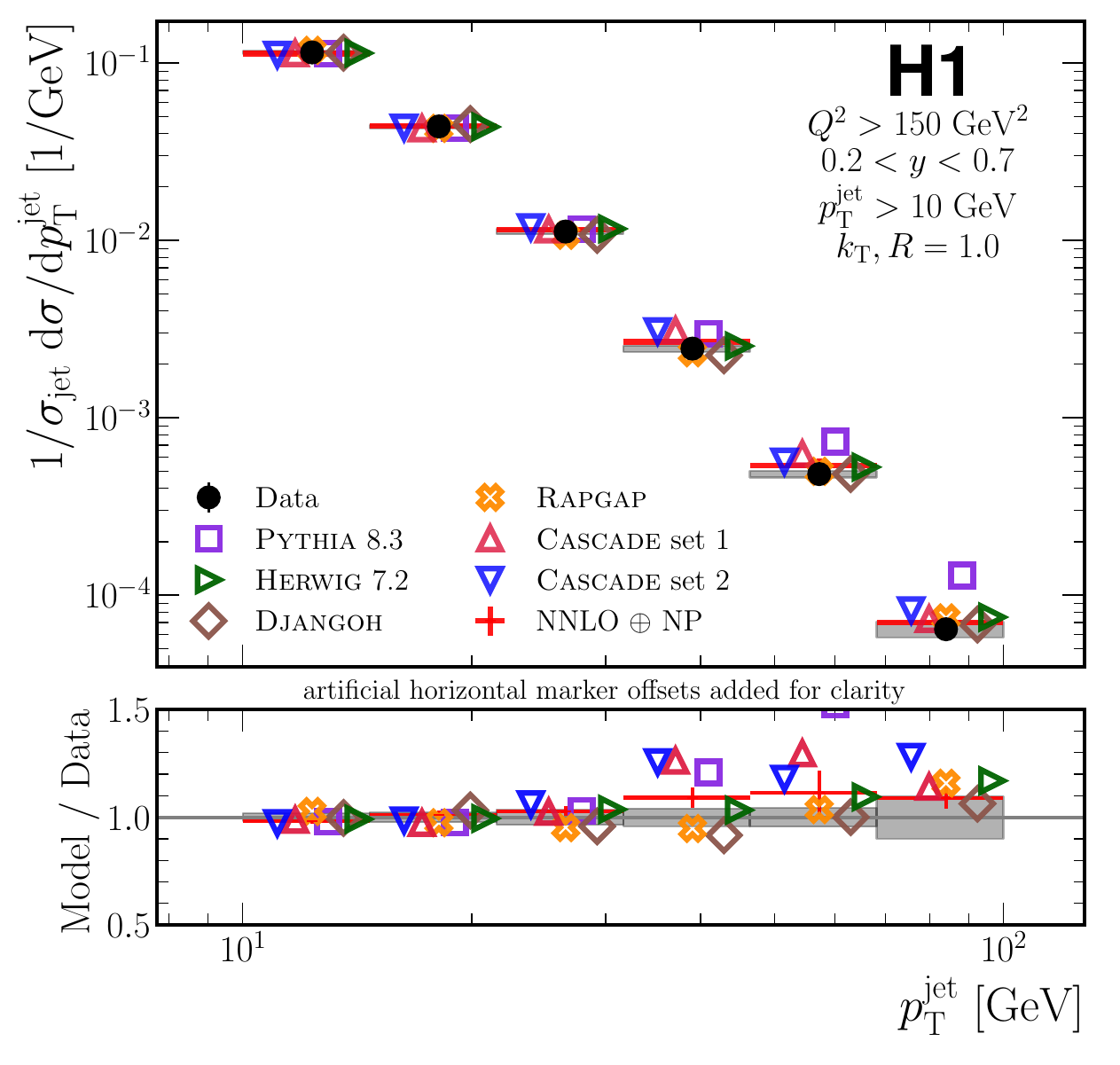}
    \includegraphics[width=0.24\textwidth]{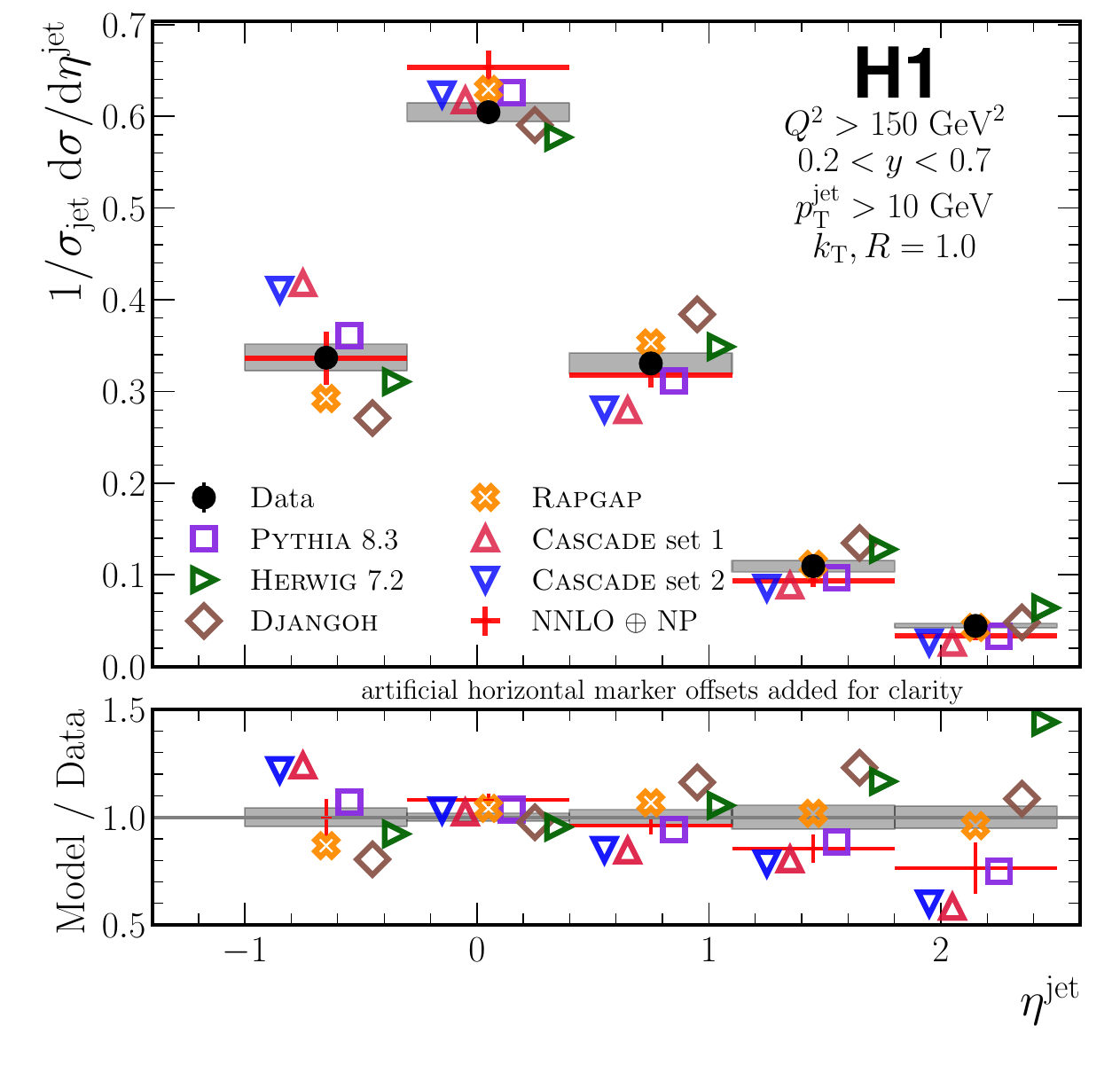}\\
    \includegraphics[width=0.24\textwidth]{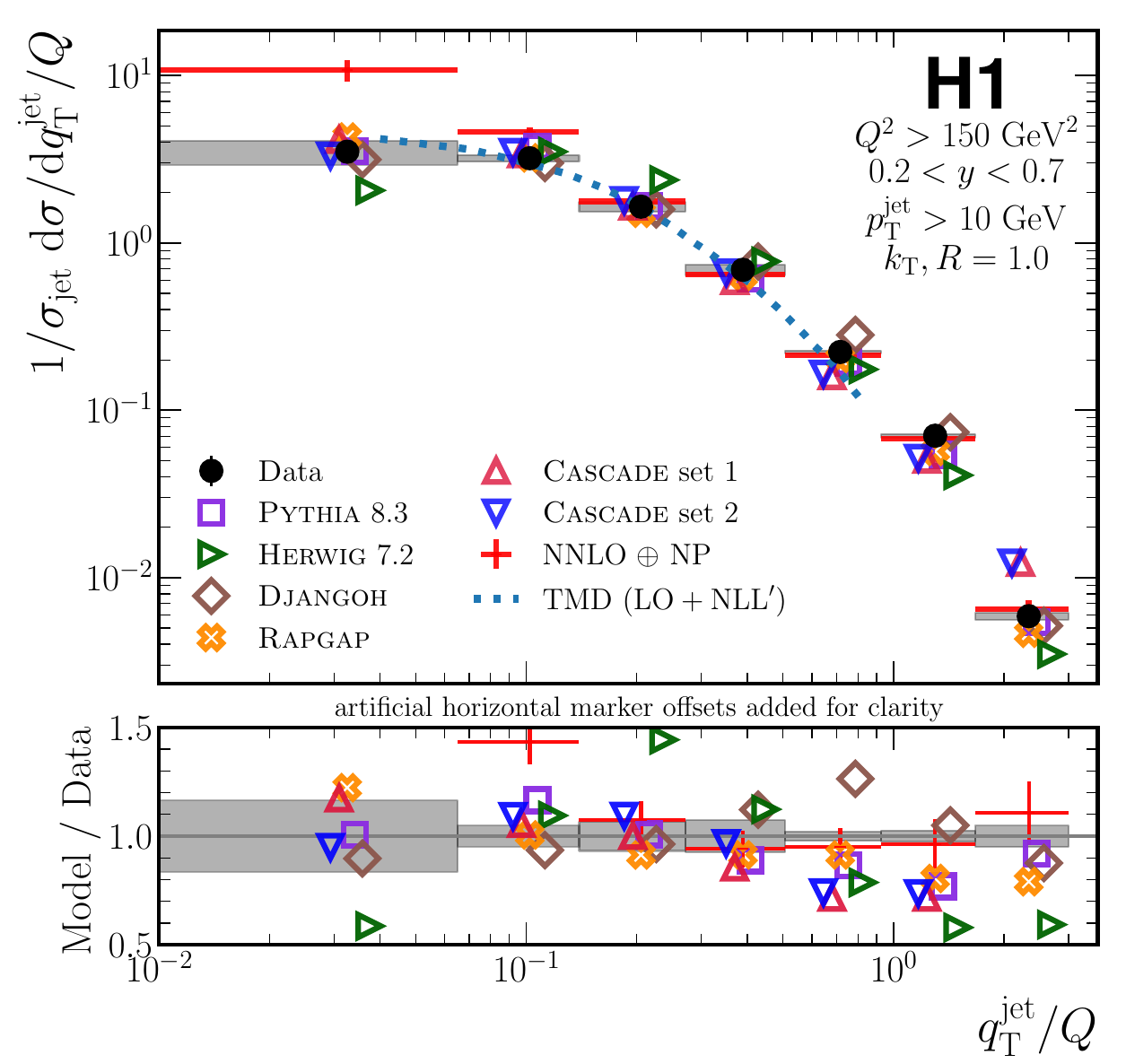}
    \includegraphics[width=0.24\textwidth]{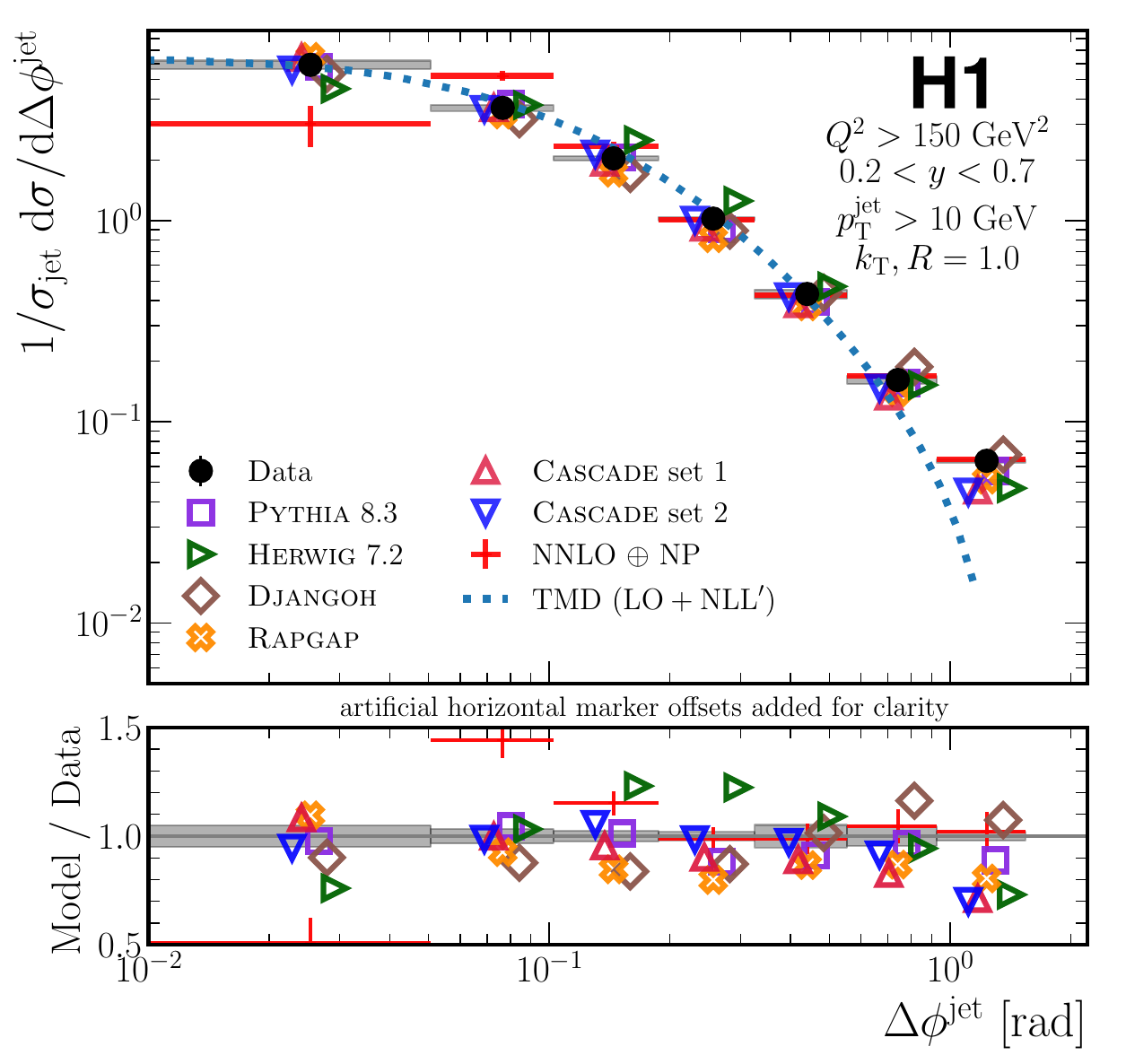}
    \caption{Measured cross sections, normalized to the inclusive jet
production cross section, as a function of the jet transverse momentum (top left) and jet pseudorapidity (top right), lepton-jet momentum balance ($q_\mathrm{T}^\text{jet}/Q$) (lower left), and lepton-jet azimuthal angle correlation ($\Delta\phi^\text{jet}$) (lower right). Predictions obtained with the pQCD (corrected by hadronization effects, ``NP'') are shown as well. Predictions obtained with the TMD framework are shown for the $q_\mathrm{T}^\text{jet}/Q$ and $\Delta\phi^\text{jet}$ cross sections. At the bottom, the ratio between predictions and the data are shown.
The gray bands represent the total systematic uncertainty of the measurement;  the bars represent the statistical uncertainty of the measurement, which is typically smaller than the marker size. The error bar on the NNLO calculation represents scale, PDF, and hadronization uncertainties.  The statistical uncertainties on the MC predictions are smaller than the markers.}
    \label{fig:Cross-sections_withtheory}
\end{figure}

\sectionPRL{Results}
The unfolded data and comparisons to predictions are presented in Fig.~\ref{fig:Cross-sections_withtheory}. The \ptjet~ and \etajetlab~cross sections are described within uncertainties by the NNLO calculation. Note that while the QED corrections are mostly small, they are up to 25\% at high \etajetlab and are essential for the observed accuracy. This result complements measurements~\cite{Chekanov:2005yb} at lower $Q^{2}$ which were found to be in good agreement with pQCD calculations~\cite{Gehrmann:2018odt}. The $q_\mathrm{T}^\text{jet}/Q$ spectrum, measured here for the first time, is described by the NNLO calculation within uncertainties in the region $q_\mathrm{T}^\text{jet}/Q>0.2$. At lower values, the predictions deviate by up to a factor of 2.5. The TMD calculation, which includes resummation, describes the data from the low $q_\mathrm{T}^\text{jet}$ to up to $q_\mathrm{T}^\text{jet}/Q\approx0.6$, which is well beyond the typically assumed validity region of the TMD framework ($q_\mathrm{T}^\text{jet}/Q\lesssim 0.25$). The agreement between the TMD calculation and data supports the underlying TMD PDFs,  soft functions, and their TMD evolution, although lack of robust theory uncertainties prevent us from drawing firm conclusions. The NNLO calculation describes the $\Delta\phi^\text{jet}$ spectrum within uncertainties, except at low $\Delta\phi^\text{jet}$ where deviations are observed, as expected since in this region soft processes dominate and contributions from logarithmic terms are enhanced. The TMD calculation describes the data well for $\Delta\phi^\text{jet} < 0.75$ rad. The overlap of the pure TMD and collinear QCD calculations over a significant region of the $q_\mathrm{T}^\text{jet}/Q$ and $\Delta\phi^\text{jet}$ spectra indicate that these data could constrain the matching between the two frameworks, which is an open problem~\cite{Collins:2016hqq}. 

\textsc{Rapgap} describes the \ptjet~and \etajetlab~cross sections within uncertainties, whereas \textsc{Djangoh} describes the \ptjet~cross section within uncertainty and shows small but significant differences with the \etajetlab~cross section. \textsc{Pythia}~8.3 describes the low \ptjet~spectrum well, but predicts a significantly harder \ptjet~spectrum beyond about 30 GeV; there are also significant deviations in the \etajetlab~cross section. \textsc{Herwig}~7.2 describes the entire \ptjet~spectrum well, but deviates from the data at high \etajetlab and for all $\Delta\phi^\text{jet}$ and $q_\text{T}^\text{jet}/Q$.  The \textsc{Cascade} calculations describe the \ptjet~spectrum well but fail for the \etajetlab~shape; they also describe the data reasonably well at low $q_\mathrm{T}^\text{jet}/Q$ and $\Delta\phi$ while missing the large values, likely due to missing higher-order contributions. While no event generator describes the $q_\mathrm{T}^\text{jet}/Q$ and $\Delta\phi^\text{jet}$ cross sections over the entire range, the data are mostly contained within the spread of predictions.

Even though uncertainties are not yet available for the TMD predictions, the spread in predictions that use different TMD sets (including \textsc{Cascade}) is comparable to the experimental and fixed-order uncertainties.  This suggests that these data will have constraining power towards a global description of TMD and collinear effects across scales.

\sectionPRL{Summary and conclusions}
Measurements of jet production in neutral current DIS events with $Q^{2}>150$ GeV$^{2}$ and $0.2<y<0.7$ have been presented. Jets are reconstructed in the laboratory frame with the $k_{\mathrm{T}}$ algorithm and distance parameter $R=1$. The following observables are measured: jet transverse momentum and pseudorapidity, as well as the TMD-sensitive observables $q_\text{T}^\text{jet}/Q$ (lepton-jet momentum imbalance) and $\Delta\phi$ (lepton-jet azimuthal angle correlation). 

This work provides the first measurement of lepton-jet imbalance at high $Q^{2}$, a variable recently proposed \cite{Liu:2018trl,Liu:2020dct} for probing TMD PDFs and their evolution. The data agree in a wide kinematic range with calculations that use TMD PDFs extracted from low $Q^{2}$ semi-inclusive DIS data and parton branching TMD PDFs extracted from other HERA data.  The experimental uncertainty is comparable to the spread from predictions using different TMD sets, suggesting that when a full TMD uncertainty breakdown is available, the data will be able to constrain the models. 

These measurements bridge the kinematic gap between DIS measurements from fixed target experiments and Drell-Yan measurements at hadron colliders, and may provide a test of TMD factorization, TMD evolution and TMD universality. These measurements complement previous and ongoing studies of TMD physics in hadronic collisions~\cite{Boer:2003tx,Abelev:2007ii,Bomhof:2007su,Adamczyk:2017wld,Kang:2017btw,DAlesio:2017bvu} and provide a baseline for jet studies in DIS of polarized protons and nuclei at the future Electron Ion Collider~\cite{Accardi:2012qut,AbdulKhalek:2021gbh}. 

This measurement also represents a milestone in the use of ML techniques for experimental physics, as it provides the first example of ML-assisted unfolding, which is based on the recently proposed \textsc{MultiFold} method~\cite{Andreassen:2019cjw} and enables simultaneous and unbinned unfolding in high dimensions. This opens up the possibility for high dimensional explorations of nucleon structure with H1 data and beyond.


\begin{acknowledgments}
\section*{Acknowledgements}

We are grateful to the HERA machine group whose outstanding efforts have made this experiment possible. We thank the engineers and technicians for their work in constructing and maintaining the H1 detector, our funding agencies for financial support, the DESY technical staff for continual assistance and the DESY directorate for support and for the hospitality which they extend to the non–DESY members of the collaboration.


We express our thanks to all those involved in securing not only the H1 data but also the software and working environment for long term use, allowing the unique H1 data set to continue to be explored. The transfer from experiment specific to central resources with long term support, including both storage and batch systems, has also been crucial to this enterprise. We therefore also acknowledge the role played by DESY-IT and all people involved during this transition and their future role in the years to come.

We thank Daniel de Florian, Ignacio Borsa and Ivan Pedron for the pQCD calculations and Feng Yuan and Zhongbo Kang for the TMD calculations, and Felix Ringer for guidance for the theory interpretation. 

\input{desy21-130.H1authorlist.revtexthanks}

\end{acknowledgments}

\raggedright

\bibliographystyle{apsrev4-2}
\bibliography{desy21-130-with-supplement.bib}

\appendix
\section*{Supplemental Material for the paper DESY 21-130}

\begin{figure}[h!]
    \centering
    \vspace*{-2.0cm}
    \includegraphics[width=0.45\textwidth]{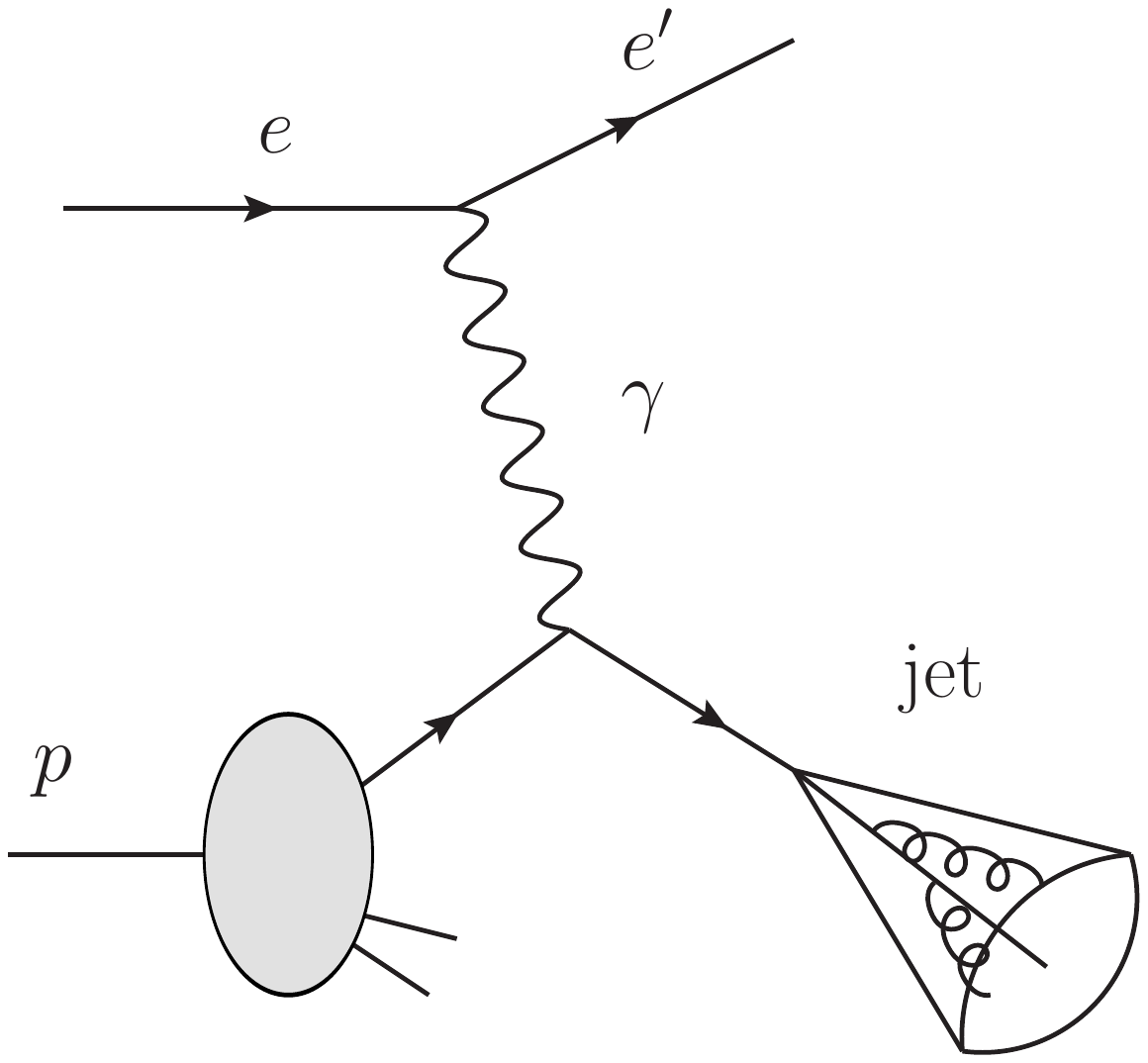}
        \vspace*{-2.0cm}
    \caption{Illustration of the neutral-current DIS.} 
    \label{fig:DISdiagram}
\end{figure}

Figure~\ref{fig:DISdiagram} illustrates the process of neutral-current deep-inelastic scattering (DIS) that is studied in this paper. The experimental signature of this reaction is shown in Fig.~\ref{fig:signature} of the main document.

\begin{figure}[h!]
    \centering
    \includegraphics[height=0.25\textheight]{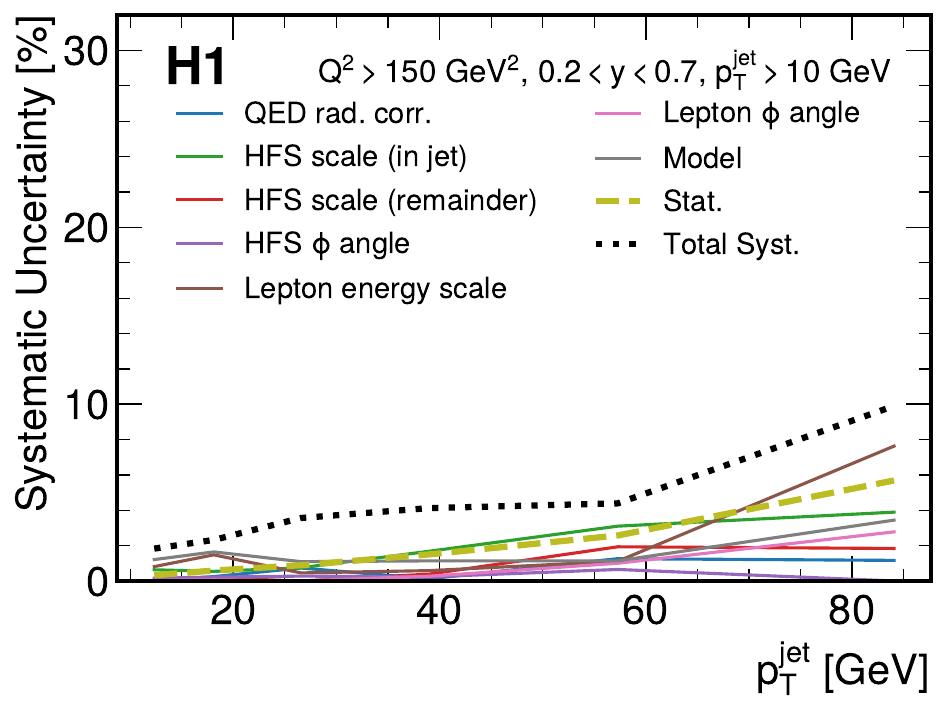}\includegraphics[height=0.25\textheight]{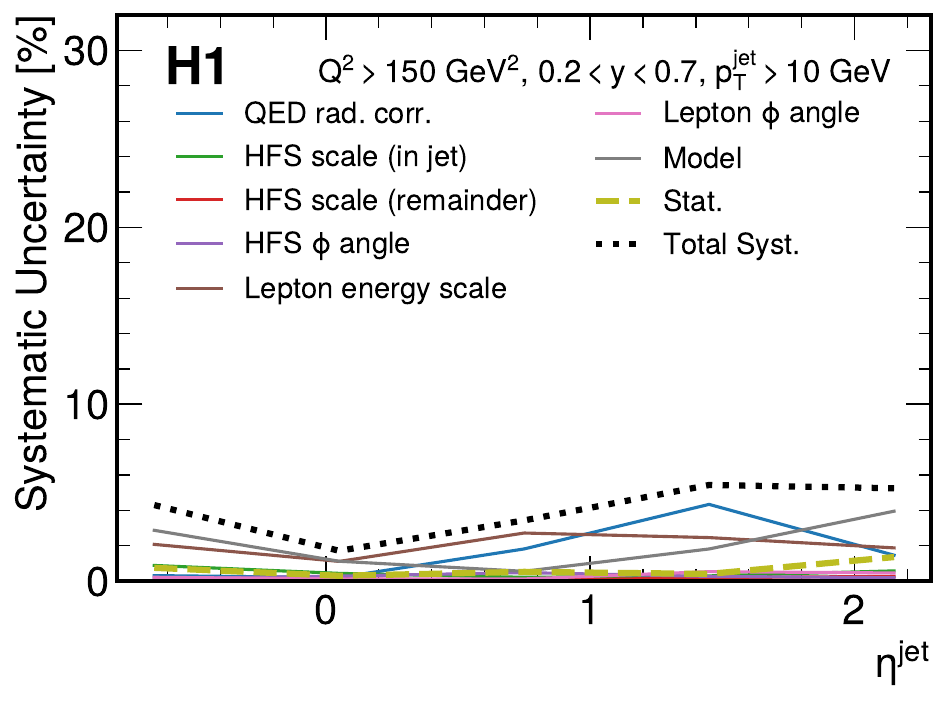}
    \includegraphics[height=0.25\textheight]{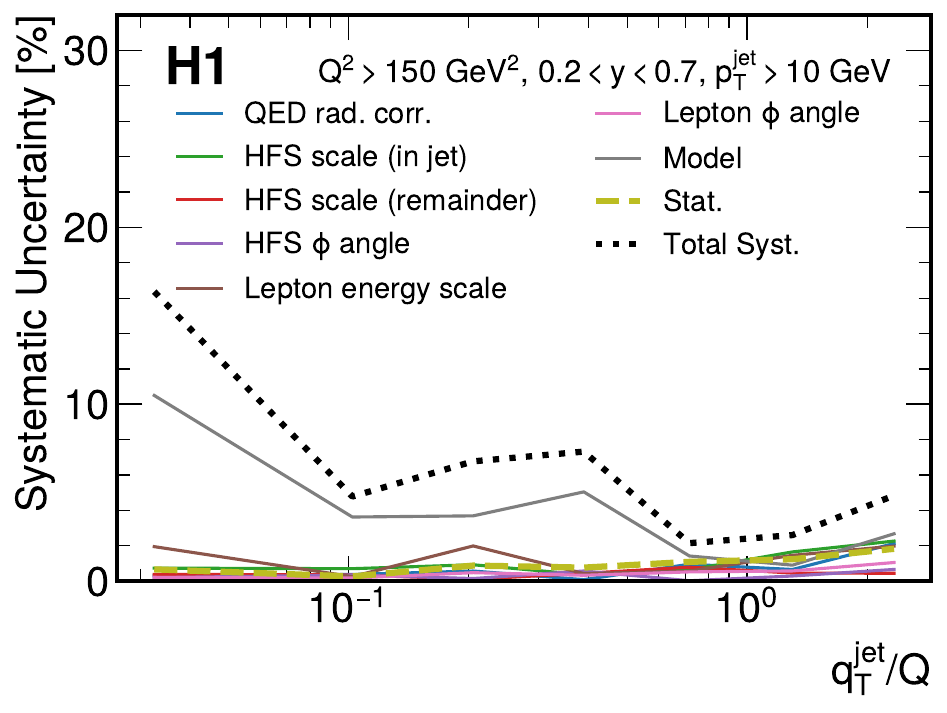}\includegraphics[height=0.25\textheight]{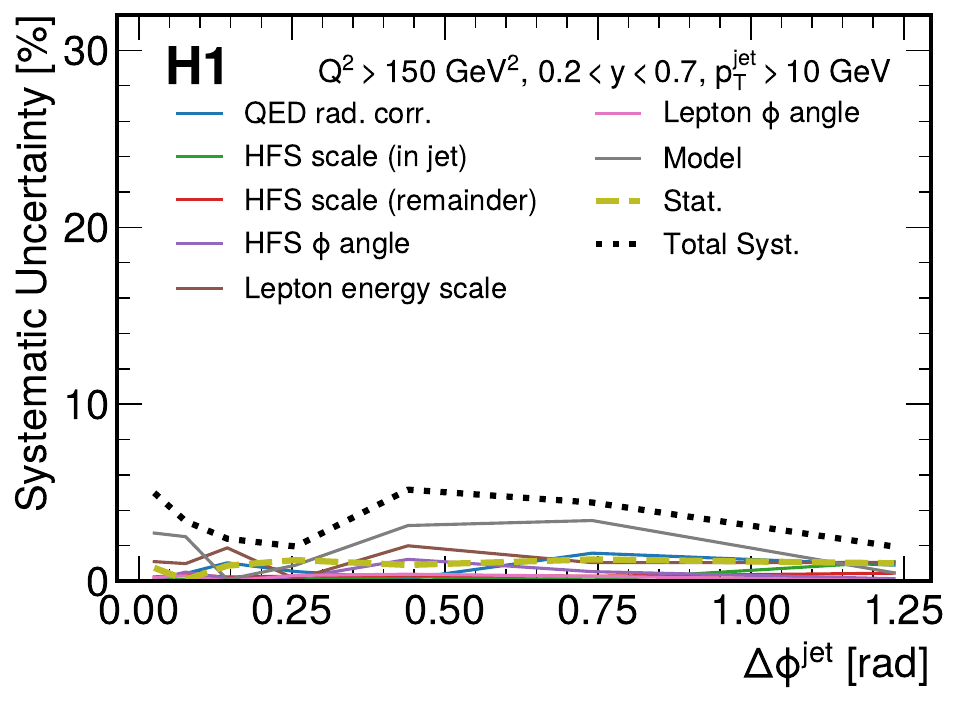}
    \caption{The uncertainty breakdown per observable.}
    \label{fig:uncerts}
\end{figure}

\begin{figure}[h!]
    \centering
    \includegraphics[height=0.6\textheight]{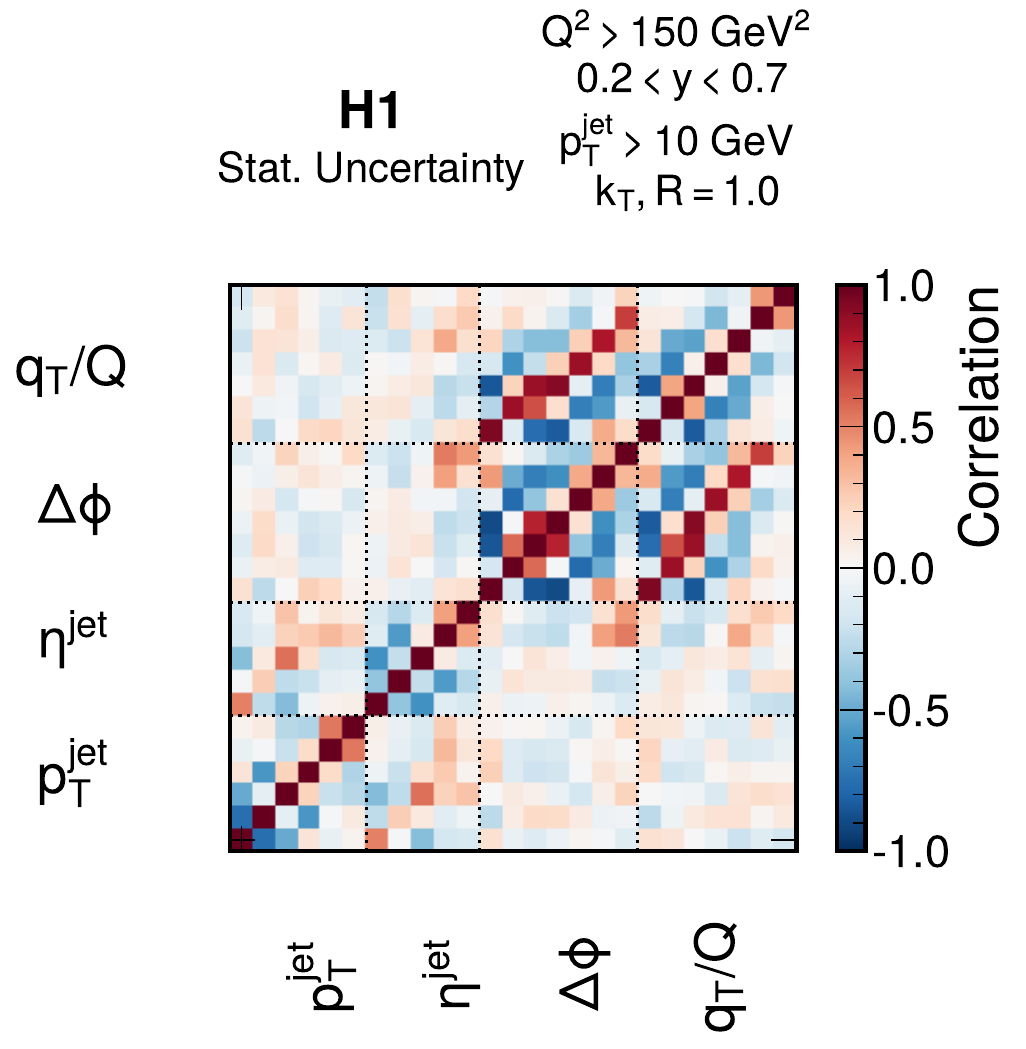}
    \caption{The statistical uncertainty correlation matrix for all measurements combined computed with 100 bootstraps of the data.}
    \label{fig:statcorall}
\end{figure}

\begin{figure}[h!]
    \centering
    \includegraphics[height=0.3\textheight]{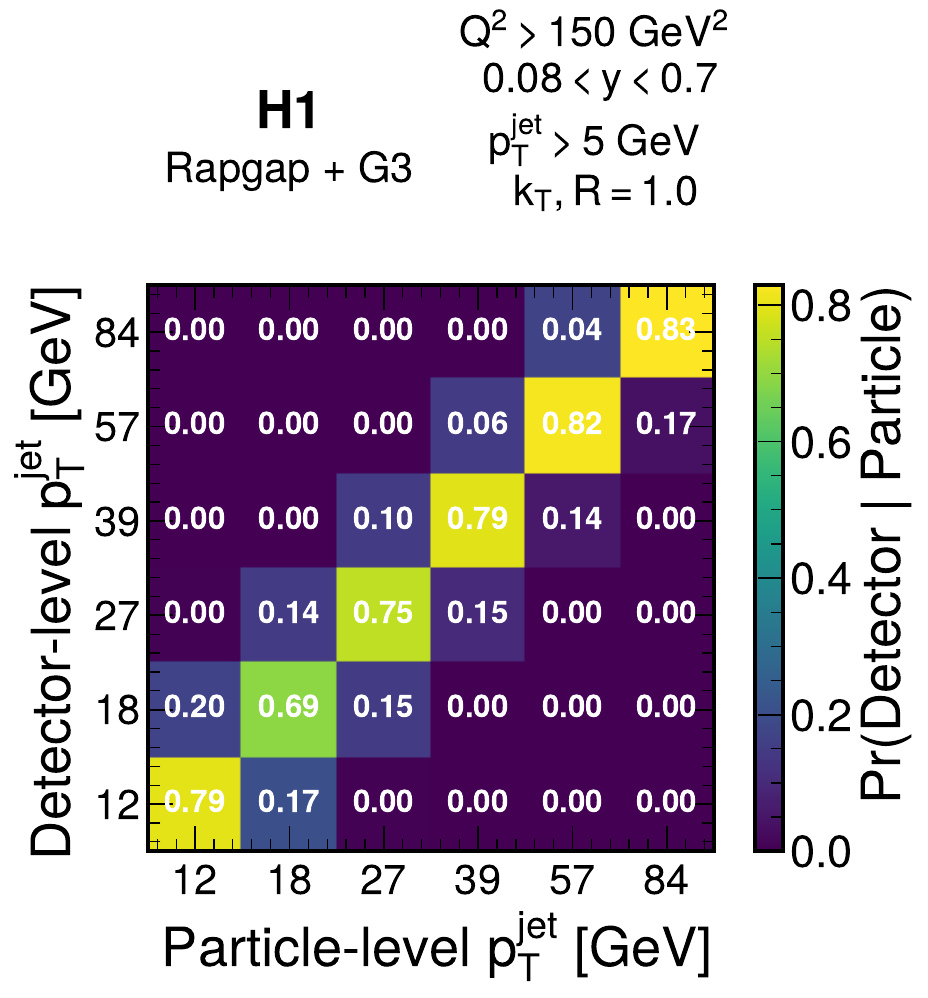}\includegraphics[height=0.3\textheight]{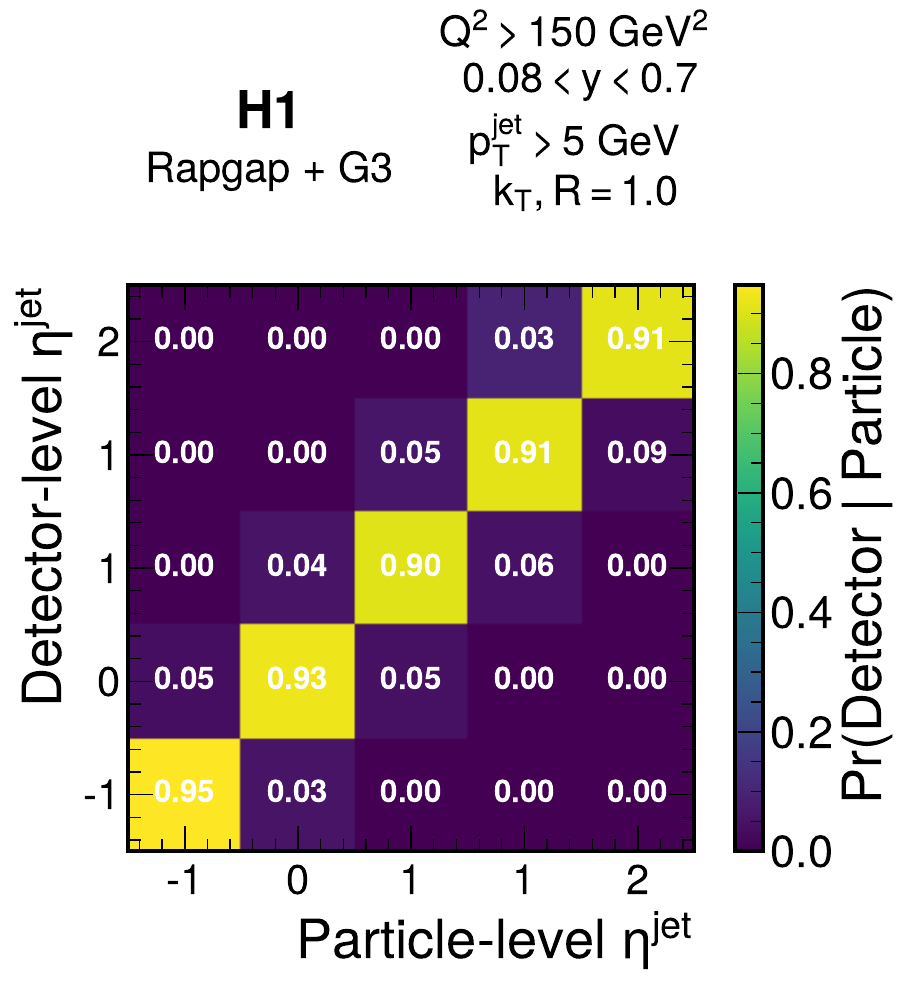}
    \includegraphics[height=0.3\textheight]{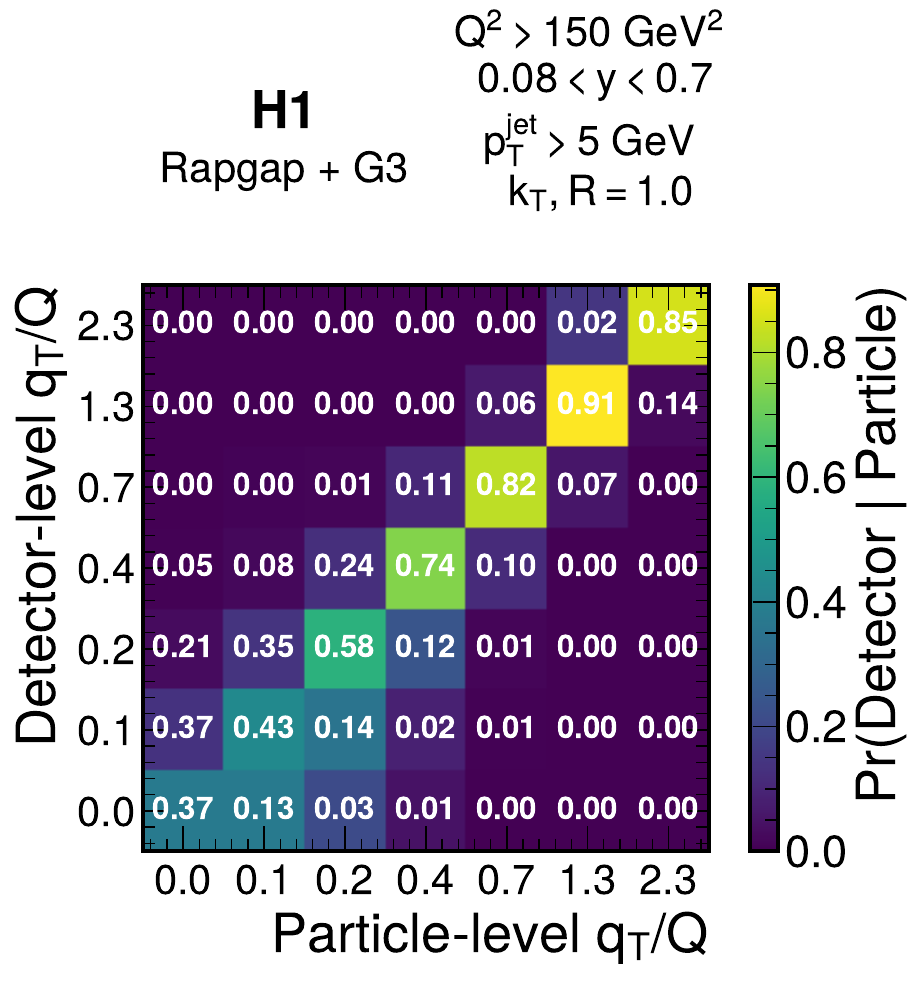}\includegraphics[height=0.3\textheight]{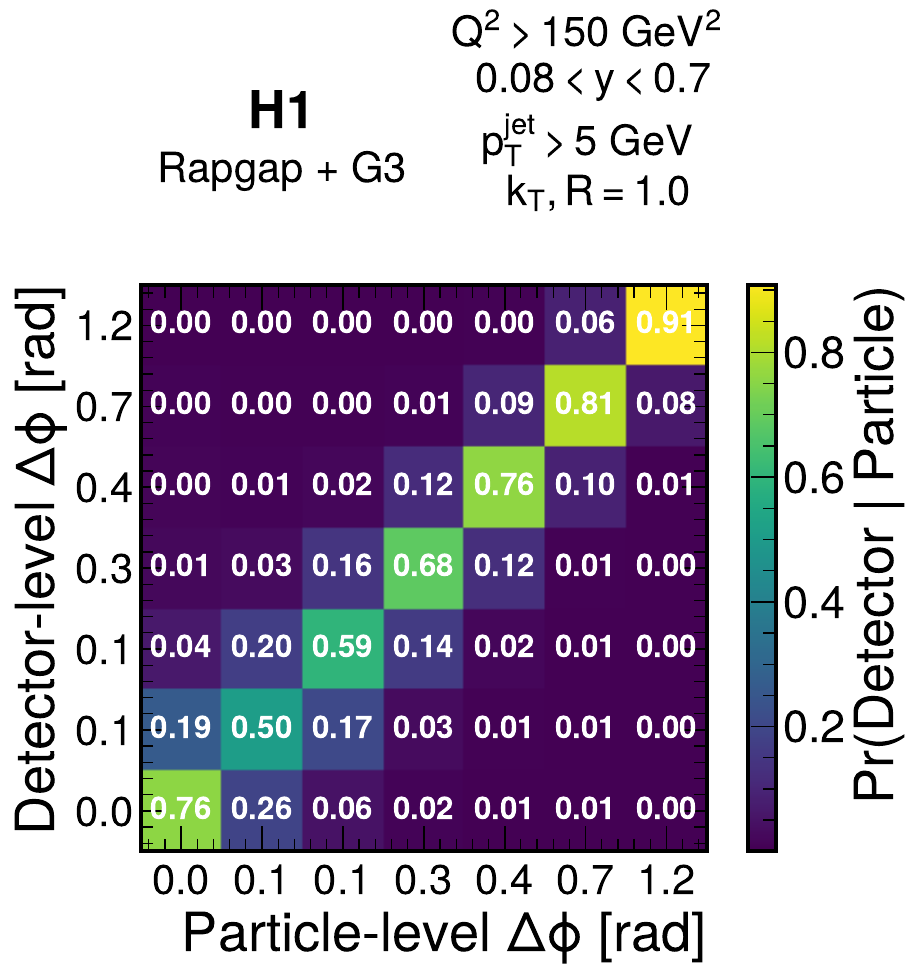}
    \caption{The response matrices per observable.  Note that these are not used in the unfolding (which is unbinned); they are shown here for illustration purposes only.}
    \label{fig:response}
\end{figure}

\begin{figure}[h!]
    \centering
    \includegraphics[height=0.25\textheight]{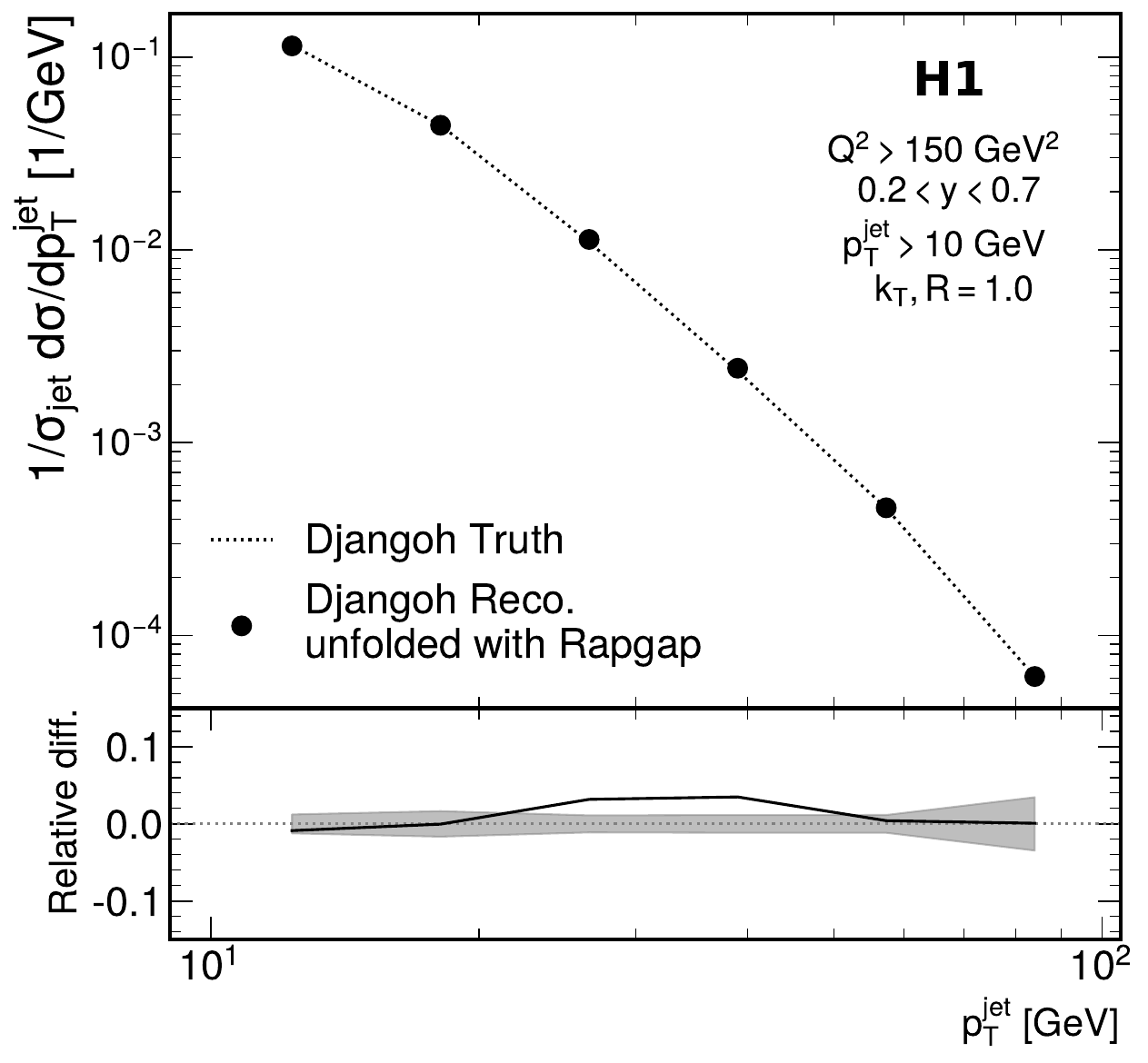}\includegraphics[height=0.25\textheight]{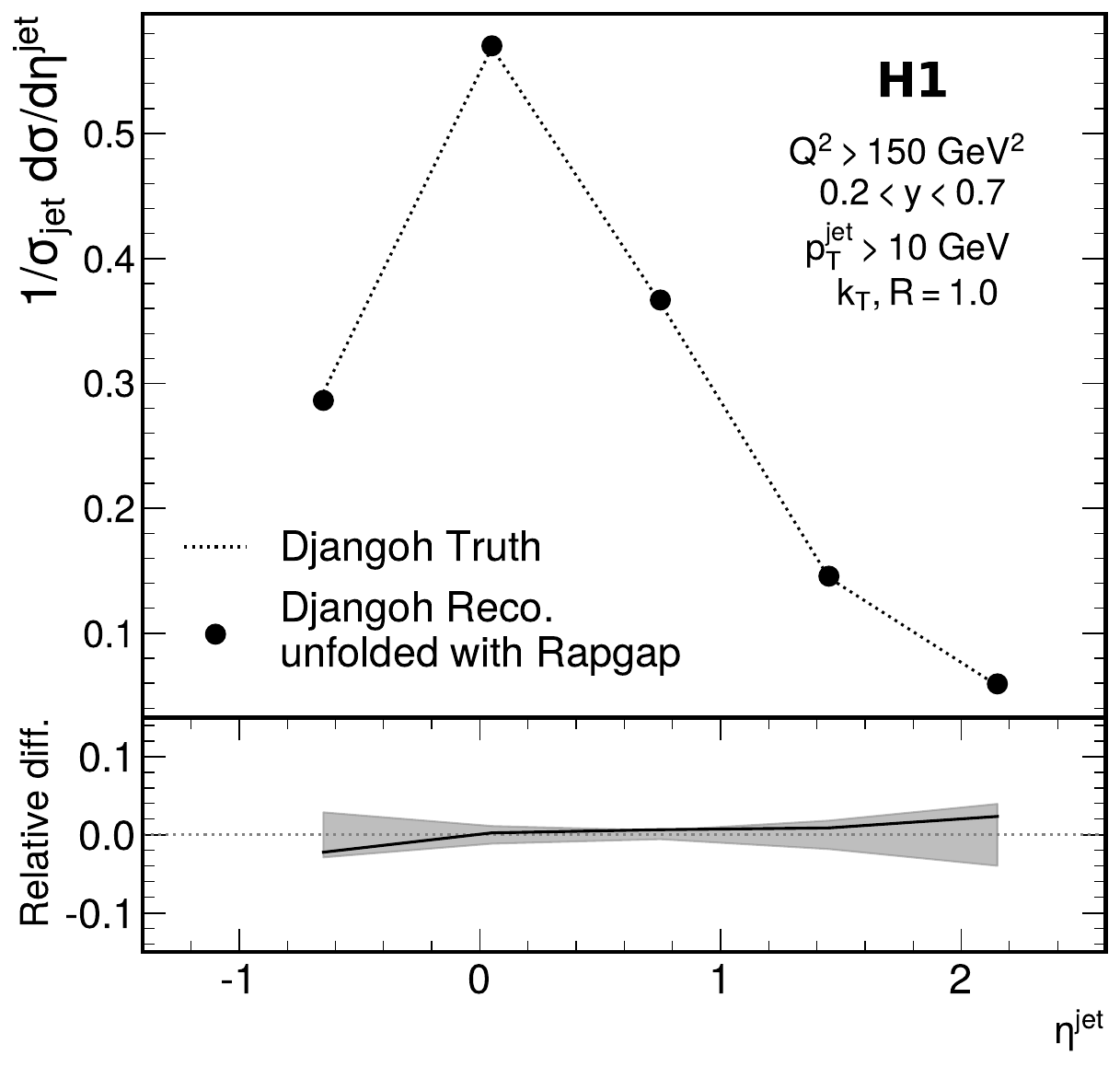}
    \includegraphics[height=0.25\textheight]{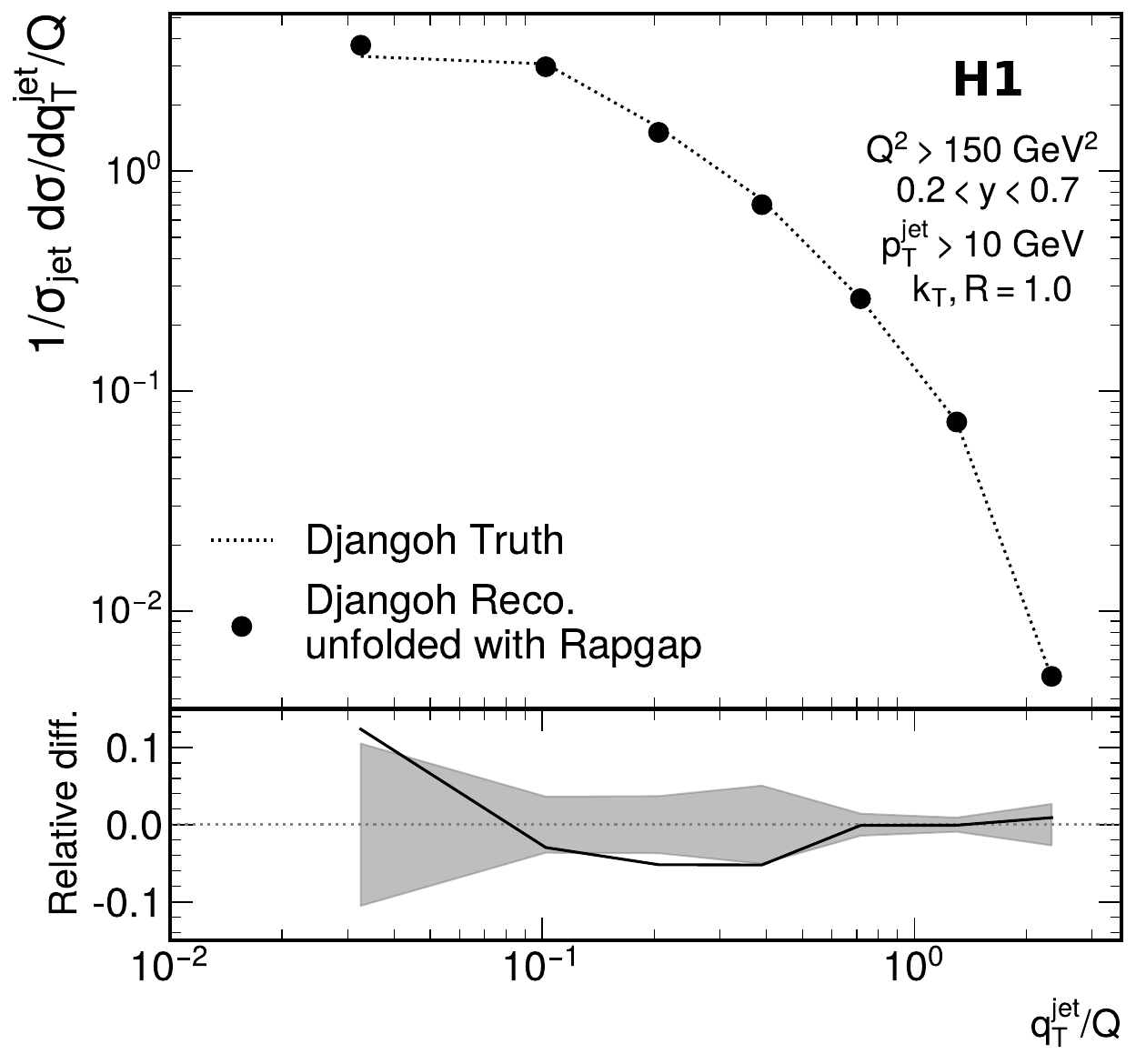}\includegraphics[height=0.25\textheight]{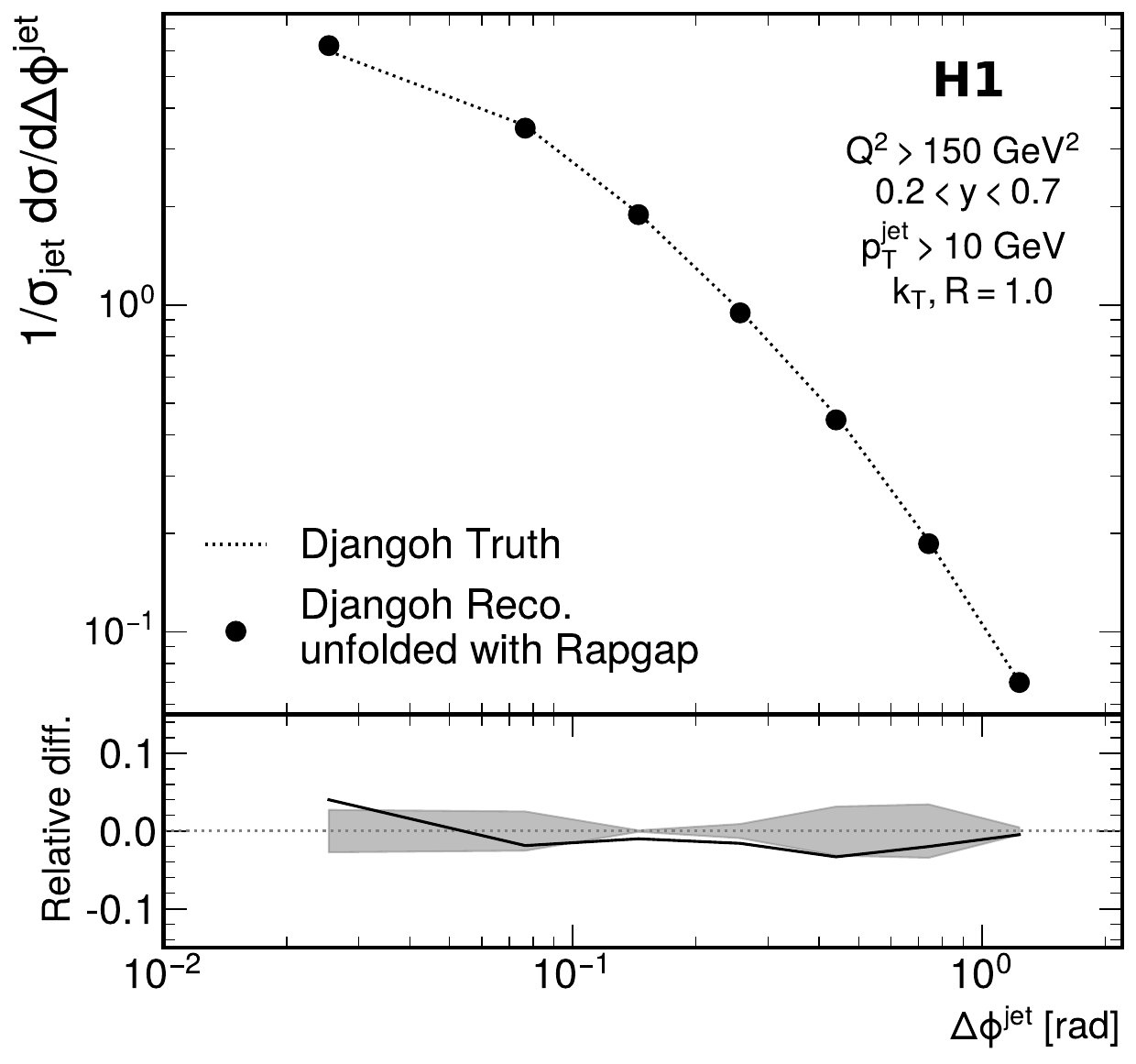}
    \caption{The method non-closure for each observable.  Detector-level Djangoh is unfolded using Rapgap (data points) and compared with the particle-level Djangoh (dashed line).  No ensembling is used for this result.  The solid line in the lower panel shows the relative difference between the unfolded and particle-level Djangoh.  The band in the lower panel is the method uncertainty reported in the measurement, computed by taking the difference between the data result unfolded by Djangoh and by Rapgap.}
    \label{fig:nonclosure}
\end{figure}

Numeric values for the measured cross sections, uncertainties, and hadronization corrections for all four observables presented in Fig.~2 of the main paper are given in Tables \ref{tab:jetpt}, \ref{tab:eta}, \ref{tab:qtq} and \ref{tab:dphi}.  The values can also be found at \url{https://www.hepdata.net}.  Note that the hadronization correction (had cor.) is not applied to the data - it is applied only to the fixed order calculations.  A graphical representation of the uncertainty breakdown can be found in Fig.~\ref{fig:uncerts}.  Systematic uncertainties of the same type are to be treated as fully correlated between observables.  The statistical correlation between bins is presented in Fig.~\ref{fig:statcorall}.  This correlation is computed by bootstrapping the data as described in the main text.  Note that there is a small contribution to the correlation from the stochastic nature of the neural network training (e.g. from random initializations) that is not subtracted. Figure~\ref{fig:response} shows response matrices per observable and the method non-closure is studied in Fig.~\ref{fig:nonclosure}.

\begin{table}[h!]\centering\begin{adjustbox}{width=1\textwidth}\begin{tabular}{| c | c | c | c || c | c | c | c | c | c | c || c | c |}\hline$p_\mathrm{T}^\mathrm{jet}$ [GeV]&$1/\sigma_\mathrm{jet}\mathrm{d}\sigma/\mathrm{d}p_\mathrm{T}^\mathrm{jet}$& $\delta_\mathrm{stat.}$& $\delta_\mathrm{tot.}$&$\delta_\mathrm{QED}$&$\delta_\mathrm{HFS(jet)}$&$\delta_\mathrm{HFS(other)}$ &$\delta_{\mathrm{HFS}(\phi)}$&$\delta_\mathrm{Lepton(E)}$&$\delta_{\mathrm{Lepton}(\phi)}$&$\delta_\mathrm{Closure}$ & had cor. & $\delta_\mathrm{had.}$\\
\hline12.3390&0.1147&0.0004&0.0021&0.0000&0.0007&0.0001&0.0002&0.0009&0.0001&0.0014&0.9808&0.0036 \\
18.1112&0.0439&0.0003&0.0010&0.0001&0.0002&0.0001&0.0001&0.0007&0.0000&0.0007&1.0314&0.0134 \\
26.5836&0.0112&0.0001&0.0004&0.0001&0.0001&0.0000&0.0000&0.0001&0.0000&0.0001&1.0123&0.0141 \\
39.01933&0.00245&0.00004&0.00010&0.00000&0.00004&0.00001&0.00001&0.00001&0.00001&0.00003&0.99659&0.04232 \\
57.27255&0.00048&0.00001&0.00002&0.00001&0.00001&0.00001&0.00000&0.00001&0.00000&0.00001&0.98674&0.10250 \\
84.064603&0.000064&0.000004&0.000006&0.000001&0.000003&0.000001&0.000000&0.000005&0.000002&0.000002&0.959525&0.013713 \\
\hline\end{tabular}\end{adjustbox}
\caption{\label{tab:jetpt}Numerical data on normalized inclusive jet cross sections 
$1/\sigma_\mathrm{jet}\mathrm{d}\sigma/\mathrm{d}p_\mathrm{T}^\mathrm{jet}$
as a function of the jet transverse momenta $p_\mathrm{T}^\mathrm{jet}$. Statistical uncertainties $\delta_\mathrm{stat.}$, total uncertainties $\delta_\mathrm{tot.}$, and the sources of systematic uncertainty $\delta_\mathrm{QED}$, 
$\delta_\mathrm{HFS(jet)}$, $\delta_\mathrm{HFS(other)}$,
$\delta_{\mathrm{HFS}(\phi)}$, $\delta_\mathrm{Lepton(E)}$,
$\delta_{\mathrm{Lepton}(\phi)}$, $\delta_\mathrm{Closure}$ are
shown. The hadronisation corrections ``had cor.'' and their
uncertainties are also given.}
\end{table}
\begin{table}[h!]\centering\begin{adjustbox}{width=1\textwidth}\begin{tabular}{| c | c | c | c || c | c | c | c | c | c | c || c | c |}\hline$\eta^\mathrm{jet}$&$1/\sigma_\mathrm{jet}\mathrm{d}\sigma/\mathrm{d}\eta^\mathrm{jet}$& $\delta_\mathrm{stat.}$& $\delta_\mathrm{tot.}$&$\delta_\mathrm{QED}$&$\delta_\mathrm{HFS(jet)}$&$\delta_\mathrm{HFS(other)}$ &$\delta_{\mathrm{HFS}(\phi)}$&$\delta_\mathrm{Lepton(E)}$&$\delta_{\mathrm{Lepton}(\phi)}$&$\delta_\mathrm{Closure}$ & had cor. & $\delta_\mathrm{had.}$\\
\hline-0.650&0.337&0.003&0.015&0.001&0.003&0.000&0.000&0.007&0.001&0.010&1.134&0.026 \\
0.050&0.605&0.002&0.010&0.001&0.003&0.001&0.002&0.007&0.001&0.007&0.993&0.014 \\
0.750&0.331&0.002&0.011&0.006&0.001&0.000&0.002&0.009&0.000&0.002&0.892&0.025 \\
1.4500&0.1096&0.0005&0.0060&0.0048&0.0003&0.0002&0.0003&0.0027&0.0006&0.0020&0.9248&0.0012 \\
2.1500&0.0444&0.0006&0.0023&0.0007&0.0003&0.0001&0.0001&0.0008&0.0002&0.0018&0.9203&0.0518 \\
\hline\end{tabular}\end{adjustbox}
\caption{\label{tab:eta}Numerical data on normalized inclusive jet cross sections
  $1/\sigma_\mathrm{jet}\mathrm{d}\sigma/\mathrm{d}\eta^\mathrm{jet}$
  as a function of the jet pseudorapidity $\eta^\mathrm{jet}$. Further
  details are specified in table \ref{tab:jetpt}.}\end{table}
\begin{table}[h!]\centering\begin{adjustbox}{width=1\textwidth}\begin{tabular}{| c | c | c | c || c | c | c | c | c | c | c || c | c |}\hline$q_\mathrm{T}^\mathrm{jet}/Q$&$1/\sigma_\mathrm{jet}\mathrm{d}\sigma/\mathrm{d}q_\mathrm{T}^\mathrm{jet}/Q$& $\delta_\mathrm{stat.}$& $\delta_\mathrm{tot.}$&$\delta_\mathrm{QED}$&$\delta_\mathrm{HFS(jet)}$&$\delta_\mathrm{HFS(other)}$ &$\delta_{\mathrm{HFS}(\phi)}$&$\delta_\mathrm{Lepton(E)}$&$\delta_{\mathrm{Lepton}(\phi)}$&$\delta_\mathrm{Closure}$ & had cor. & $\delta_\mathrm{had.}$\\
\hline0.03&3.51&0.02&0.57&0.01&0.03&0.01&0.01&0.07&0.01&0.37&0.99&0.06 \\
0.102&3.207&0.009&0.154&0.012&0.023&0.012&0.011&0.008&0.006&0.116&0.958&0.052 \\
0.21&1.65&0.01&0.11&0.01&0.02&0.00&0.00&0.03&0.01&0.06&0.99&0.06 \\
0.389&0.691&0.005&0.051&0.001&0.003&0.003&0.004&0.003&0.002&0.035&1.047&0.060 \\
0.716&0.223&0.002&0.005&0.002&0.001&0.002&0.000&0.001&0.001&0.003&1.076&0.020 \\
1.2988&0.0705&0.0009&0.0018&0.0005&0.0012&0.0003&0.0002&0.0010&0.0004&0.0006&1.0647&0.0139 \\
2.3359&0.0059&0.0001&0.0003&0.0001&0.0001&0.0000&0.0000&0.0001&0.0001&0.0002&1.0934&0.0459 \\
\hline\end{tabular}\end{adjustbox}
\caption{\label{tab:qtq}Numerical data on normalized inclusive jet cross sections
$1/\sigma_\mathrm{jet}\mathrm{d}\sigma/\mathrm{d}q_\mathrm{T}^\mathrm{jet}/Q$
as a function of the scaled lepton-jet relative transverse momenta
$q_\mathrm{T}^\mathrm{jet}/Q$. The relative momenta $q_\mathrm{T}$ are
scaled by the momentum transfer $Q$ as explained in the main text. 
Further details are specified in table \ref{tab:jetpt}.}
\end{table}
\begin{table}[h!]\centering\begin{adjustbox}{width=1\textwidth}\begin{tabular}{| c | c | c | c || c | c | c | c | c | c | c || c | c |}\hline$\Delta\phi^\mathrm{jet}$ &$1/\sigma_\mathrm{jet}\mathrm{d}\sigma/\mathrm{d}\Delta\phi^\mathrm{jet}$ & $\delta_\mathrm{stat.}$& $\delta_\mathrm{tot.}$&$\delta_\mathrm{QED}$&$\delta_\mathrm{HFS(jet)}$&$\delta_\mathrm{HFS(other)}$ &$\delta_{\mathrm{HFS}(\phi)}$&$\delta_\mathrm{Lepton(E)}$&$\delta_{\mathrm{Lepton}(\phi)}$&$\delta_\mathrm{Closure}$ & had cor. & $\delta_\mathrm{had.}$\\
\hline0.03&5.93&0.05&0.30&0.00&0.00&0.01&0.01&0.07&0.01&0.16&0.98&0.01 \\
0.077&3.622&0.003&0.123&0.016&0.004&0.013&0.019&0.036&0.011&0.091&0.973&0.030 \\
0.14&2.03&0.02&0.05&0.02&0.00&0.00&0.00&0.04&0.00&0.00&0.98&0.02 \\
0.26&1.02&0.01&0.02&0.01&0.00&0.00&0.00&0.00&0.00&0.01&1.02&0.00 \\
0.440&0.431&0.004&0.022&0.001&0.001&0.001&0.005&0.009&0.002&0.014&1.053&0.029 \\
0.741&0.161&0.002&0.007&0.003&0.000&0.000&0.001&0.002&0.000&0.006&1.074&0.004 \\
1.2343&0.0640&0.0007&0.0013&0.0006&0.0007&0.0003&0.0001&0.0007&0.0000&0.0003&1.0594&0.0139 \\
\hline\end{tabular}\end{adjustbox}
\caption{\label{tab:dphi}Numerical data on normalized inclusive jet cross sections
$1/\sigma_\mathrm{jet}\mathrm{d}\sigma/\mathrm{d}\Delta\phi^\mathrm{jet}$
as a function of the lepton-jet azimuthal angular difference $\Delta\phi^\mathrm{jet}$.
Further details are specified in table \ref{tab:jetpt}.}
\end{table}

\clearpage

\section{Brief Review of MultiFold}

This section briefly reviews the \textsc{MultiFold} technique introduced in Ref.~\cite{Andreassen:2019cjw,Andreassen:2021zzk}.  Let $\vec{x}=(\vec{p}_\mathrm{T}^e, p_z^e, p_\mathrm{T}^\text{jet}, \eta^\text{jet}, \phi^\text{jet}, q_\mathrm{T}^\text{jet}/Q, \Delta\phi^\text{jet})$. \textsc{MultiFold} is an iterative, two-step procedure.  Let $X_\text{data}=\{\vec{x}_i\}$ be the set of events in data and $X_\text{MC,truth}=\{\vec{x}_\text{truth,$i$}\}$ and $X_\text{MC,reco}=\{\vec{x}_\text{reco,$i$}\}$ be sets of events in simulation with a correspondance between the two sets.  In simulation, we have a set of observables at particle-level (`truth') and detector-level (`reco') for each event.  If an event does not pass the particle-level or detector-level event selection, then the corresponding set of observables are assigned a dummy value $\vec{x}=\emptyset$.  Each event $i$ in simulation is also associated with a weight $w_i$.  \textsc{MultiFold} then proceeds iteratively by repeating the following two steps to iteratively adjust a set of event weights $\nu_i$:

\begin{enumerate}
    \item Train a classifier $f$ to distinguish the weighted simulation at detector-level from the data.  The loss function is the binary cross entropy:
    
    \begin{align}
        L_1[f]=-\sum_{\vec{x}_i\in X_\text{data}}\log(f(\vec{x}_i)) - \sum_{\vec{x}_i\in X_\text{MC,reco}}\nu_i\,w_i\, \log(1-f(\vec{x}_i))\,,
    \end{align}
    
    where both sums only include events that pass the detector-level selection.  For events that pass the detector-level selection, define $\lambda_i=\nu_i\times f(\vec{x}_i)/(1-f(\vec{x}_i))$ for $\vec{x}_i\in X_\text{MC,reco}$.  For events that do not pass the detector-level selection, $\lambda_i=\nu_i$.
    
    \item Train a classifier $g$ to distinguish the particle-level simulation weighted by $\nu$ from the particle-level simulation weighted by $\lambda$.  The loss function is once again the binary cross entropy:
    
    \begin{align}
        L_2[g]=-\sum_{\vec{x}_i\in X_\text{MC,truth}}\lambda_i\,w_i\, \log(g(\vec{x}_i)) - \sum_{\vec{x}_i\in X_\text{MC,truth}}\nu_i\,w_i\,\log(1-g(\vec{x}_i))\,,
    \end{align}
    
     where both sums only include events that pass the particle-level selection. For events that pass the particle-level selection, define $\nu_i=\nu_i\times g(\vec{x}_i)/(1-g(\vec{x}_i))$ for $\vec{x}_i\in X_\text{MC,truth}$.  For events that do not pass the particle-level selection, $\nu_i$ is left unchanged from its previous value.
    
\end{enumerate}

\noindent The process is initialized by $\nu_i=1$ for all events. The $f/(1-f)$ or $g/(1-g)$ form for the weights is a well-known (see e.g. Ref.~\cite{hastie01statisticallearning,sugiyama_suzuki_kanamori_2012}) approximation for the likelihood ratio of the two samples in the left and right sums in each equation.  After iterating the above procedure some number of times, the final result is constructed by making histograms with the truth events using the final $\{\nu_i w_i\}$ weights.

\section{Neural Network Settings}

Neural networks were trained using three computing systems: one with NVDIA A40 graphical processing units (GPUs), \texttt{python} 3.8.8, \texttt{tensorflow} 2.5.0, and \texttt{numpy} 1.19.5; one with NVIDIA RTX 6000 GPUs, \texttt{python} 3.8.5, \texttt{tensorflow} 2.2.0, and \texttt{numpy} 1.19.2; and one with NVIDIA V100 GPUs, \texttt{python} 3.9.4, \texttt{tensorflow} 2.4.1, and \texttt{numpy} 1.20.1.  All neural networks were composed of three hidden layers with 50 nodes in the first layer, 100 nodes in the second layer, and 50 nodes in the last intermediate layer.  Rectified linear unit activation functions are used for all intermediate layers and a sigmoid activation is used in the last layer.  None of these hyperparameters were optimized and all other hyperparemeters are set to their default values.  In particular, the biases are all initialized to zero and the weights are initialized using the Glorot uniform distribution~\cite{pmlr-v9-glorot10a}.


\end{document}

%% file: desy21-130.H1authorlist.revtexnothanks.tex
\affiliation{I. Physikalisches Institut der RWTH, Aachen, Germany}
\affiliation{LAPP, Université de Savoie, CNRS/IN2P3, Annecy-le-Vieux, France}
\affiliation{University of Michigan, Ann Arbor, MI 48109, USA$^{f1}$}
\affiliation{Inter-University Institute for High Energies ULB-VUB, Brussels and Universiteit Antwerpen, Antwerp, Belgium$^{f2}$}
\affiliation{Lawrence Berkeley National Laboratory, Berkeley, CA 94720, USA$^{f1}$}
\affiliation{School of Physics and Astronomy, University of Birmingham, Birmingham, United Kingdom$^{f3}$}
\affiliation{Horia Hulubei National Institute for R\&D in Physics and Nuclear Engineering (IFIN-HH) , Bucharest, Romania$^{f4}$}
\affiliation{STFC, Rutherford Appleton Laboratory, Didcot, Oxfordshire, United Kingdom$^{f3}$}
\affiliation{Institut für Physik, TU Dortmund, Dortmund, Germany$^{f5}$}
\affiliation{Joint Institute for Nuclear Research, Dubna, Russia}
\affiliation{CERN, Geneva, Switzerland}
\affiliation{Irfu/SPP, CE Saclay, Gif-sur-Yvette, France}
\affiliation{II. Physikalisches Institut, Universität Göttingen, Göttingen, Germany}
\affiliation{Deutsches Elektronen-Synchrotron DESY, Hamburg, Germany}
\affiliation{Physikalisches Institut, Universität Heidelberg, Heidelberg, Germany$^{f5}$}
\affiliation{Rice University, Houston, TX 77005-1827, USA}
\affiliation{Institute of Nuclear Physics Polish Academy of Sciences, Krakow, Poland$^{f6}$}
\affiliation{Department of Physics, University of Lancaster, Lancaster, United Kingdom$^{f3}$}
\affiliation{Department of Physics, University of Liverpool, Liverpool, United Kingdom$^{f3}$}
\affiliation{School of Physics and Astronomy, Queen Mary, University of London, London, United Kingdom$^{f3}$}
\affiliation{Aix Marseille Univ, CNRS/IN2P3, CPPM, Marseille, France}
\affiliation{Institute for Theoretical and Experimental Physics, Moscow, Russia$^{f7}$}
\affiliation{Lebedev Physical Institute, Moscow, Russia}
\affiliation{Lomonosov Moscow State University, Skobeltsyn Institute of Nuclear Physics, Moscow, Russia}
\affiliation{Institute for Information Transmission Problems RAS, Moscow, Russia$^{f8}$}
\affiliation{Max-Planck-Institut für Physik, München, Germany}
\affiliation{Oak Ridge National Laboratory, Oak Ridge, TN 37831, USA}
\affiliation{Joint Laboratory of Optics, Palack\`y University, Olomouc, Czech Republic}
\affiliation{IJCLab, Université Paris-Saclay, CNRS/IN2P3, Orsay, France}
\affiliation{Department of Physics, Oxford University, Oxford, United Kingdom}
\affiliation{LLR, Ecole Polytechnique, CNRS/IN2P3, Palaiseau, France}
\affiliation{Faculty of Science, University of Montenegro, Podgorica, Montenegro$^{f9}$}
\affiliation{Institute of Physics, Academy of Sciences of the Czech Republic, Praha, Czech Republic$^{f10}$}
\affiliation{Faculty of Mathematics and Physics, Charles University, Praha, Czech Republic$^{f10}$}
\affiliation{University of California, Riverside, CA 92521, USA}
\affiliation{Dipartimento di Fisica Università di Roma Tre and INFN Roma 3, Roma, Italy}
\affiliation{Shandong University, Shandong, P.R.China}
\affiliation{Stony Brook University, Stony Brook, NY 11794, USA$^{f1}$}
\affiliation{Institute of Physics and Technology of the Mongolian Academy of Sciences, Ulaanbaatar, Mongolia}
\affiliation{Ulaanbaatar University, Ulaanbaatar, Mongolia}
\affiliation{Brookhaven National Laboratory, Upton, NY 11973, USA}
\affiliation{Université Claude Bernard Lyon 1, CNRS/IN2P3, Villeurbanne, France}
\affiliation{Paul Scherrer Institut, Villigen, Switzerland}
\affiliation{Department of Physics and Astronomy, Purdue University, West Lafayette, IN 47907, USA}
\affiliation{Fachbereich C, Universität Wuppertal, Wuppertal, Germany}
\affiliation{Yerevan Physics Institute, Yerevan, Armenia}
\affiliation{Departamento de Fisica Aplicada, CINVESTAV, Mérida, Yucat\'an, México$^{f11}$}
\affiliation{Deutsches Elektronen-Synchrotron DESY, Zeuthen, Germany}
\affiliation{Institut für Teilchenphysik, ETH, Zürich, Switzerland$^{f12}$}
\affiliation{Physik-Institut der Universität Zürich, Zürich, Switzerland$^{f12}$}
\author{V.~Andreev}
\affiliation{Lebedev Physical Institute, Moscow, Russia}
\author{M.~Arratia}
\affiliation{University of California, Riverside, CA 92521, USA}
\author{A.~Baghdasaryan}
\affiliation{Yerevan Physics Institute, Yerevan, Armenia}
\author{A.~Baty}
\affiliation{Rice University, Houston, TX 77005-1827, USA}
\author{K.~Begzsuren}
\affiliation{Institute of Physics and Technology of the Mongolian Academy of Sciences, Ulaanbaatar, Mongolia}
\author{A.~Belousov}
\thanks{deceased}
\affiliation{Lebedev Physical Institute, Moscow, Russia}
\author{A.~Bolz}
\affiliation{Deutsches Elektronen-Synchrotron DESY, Hamburg, Germany}
\author{V.~Boudry}
\affiliation{LLR, Ecole Polytechnique, CNRS/IN2P3, Palaiseau, France}
\author{G.~Brandt}
\affiliation{II. Physikalisches Institut, Universität Göttingen, Göttingen, Germany}
\author{D.~Britzger}
\affiliation{Max-Planck-Institut für Physik, München, Germany}
% ...
\collaboration{The H1 Collaboration}\noaffiliation

%% file: desy21-130.H1authorlist.revtexthanks.tex
\par $^{f1}$ supported by the U.S. DOE Office of Science
\par $^{f2}$ supported by FNRS-FWO-Vlaanderen, IISN-IIKW and IWT and by Interuniversity Attraction Poles Programme, Belgian Science Policy
\par $^{f3}$ supported by the UK Science and Technology Facilities Council, and formerly by the UK Particle Physics and Astronomy Research Council
\par $^{f4}$ supported by the Romanian National Authority for Scientific Research under the contract PN 09370101
\par $^{f5}$ supported by the Bundesministerium für Bildung und Forschung, FRG, under contract numbers 05H09GUF, 05H09VHC, 05H09VHF, 05H16PEA
\par $^{f6}$ partially supported by Polish Ministry of Science and Higher Education, grant DPN/N168/DESY/2009
\par $^{f7}$ Russian Foundation for Basic Research (RFBR), grant no 1329.2008.2 and Rosatom
\par $^{f8}$ Russian Foundation for Sciences, project no 14-50-00150
\par $^{f9}$ partially supported by Ministry of Science of Montenegro, no. 05-1/3-3352
\par $^{f10}$ supported by the Ministry of Education of the Czech Republic under the project INGO-LG14033
\par $^{f11}$ supported by CONACYT, México, grant 48778-F
\par $^{f12}$ supported by the Swiss National Science Foundation